\documentclass[aps,prd,amsmath,floats,floatfix,showpacs,nofootinbib,10pt,longbibliography]{revtex4-2}
 
\usepackage{amssymb}
\usepackage{amsmath}
\usepackage{mathtools}
\usepackage{verbatim}
\usepackage{mathrsfs}
\usepackage{amsfonts}
\usepackage{latexsym}
\usepackage{epsfig}
\usepackage{color}
\usepackage{graphicx}
\usepackage{units}
\usepackage{overpic}
\usepackage[hidelinks]{hyperref}
\usepackage{arydshln}
\usepackage{afterpage}
\usepackage{float}
\usepackage{yhmath}

\usepackage{xcolor}
\hypersetup{
    colorlinks,
    linkcolor={red!50!black},
    citecolor={blue!50!black},
    urlcolor={blue!80!black}
}

\usepackage{soulpos}
\usepackage{xcolor}
\ulposdef{\hlst}{
    \rlap{\textcolor{yellow}{\rule[-.75ex]{\ulwidth}{2.5ex}}}%
    \rule[.45ex]{\ulwidth}{.1ex}%
} 

\usepackage{placeins}

\newcommand{\figlen}{0.45}

\begin{document}


\definecolor{orange}{rgb}{0.9,0.45,0}

\newcommand{\mm}[1]{\textcolor{orange}{{\bf #1}}}



\title{\texorpdfstring{$\ell$}{l}-Boson stars in anti-de Sitter spacetime}

\author{Miguel Megevand}
\email{mfmegevand@gmail.com}
\affiliation{Instituto de F\'{\i}sica Enrique Gaviola, CONICET. Ciudad Universitaria, 5000 C\'ordoba, Argentina}


\date{\today}


\begin{abstract}
  In previous work, we introduced the $\ell$-boson stars, a generalization of standard boson stars, which are parameterized by an angular momentum number $\ell$, while still preserving the spacetime's spherical symmetry.
  In this article, we present and study the properties of $\ell$-boson stars in spacetimes with a negative cosmological constant, such that they are asymptotically anti-de~Sitter.
\end{abstract}




\maketitle



\section{Introduction}  \label{Sec:Intro}  
First introduced by Kaup~\cite{Kaup:1968zz} and by Ruffini and Bonazzola~\cite{Ruffini:1969qy} in the late sixties, and studied in more detail in the following decades~\cite{Colpi:1986ye, Friedberg:1986tp, Gleiser:1988rq, Lee:1988av, Seidel:1990jh}, boson stars~\cite{Liddle:1992fmk, Jetzer:1991jr, Mielke:1997re, Schunck:2003kk, Mielke:2016war, Visinelli:2021uve, Shnir:2022lba, Liebling:2012fv} have more recently been of new interest for several physical applications.  
In the context of galaxies, they can provide a good model for the dark matter core~\cite{Sin:1992bg, Schive:2014dra}, even after accounting for a central super-massive black hole~\cite{Alcubierre:2024mtq, Alcubierre:2025zus}. 
Another motivation is that they can model axion stars~\cite{Visinelli:2017ooc, Braaten:2015eeu, Guerra:2019srj}, formed by axions~\cite{Peccei:1977hh, Wilczek:1977pj, Weinberg:1977ma} or axion-like particles, which have also been proposed as dark matter candidates~\cite{Sikivie:2024isv}. 
Furthermore, boson stars have been considered as plausible black hole mimickers~\cite{Torres:2000dw, Guzman:2005bs, Amaro-Seoane:2010pks, Cardoso:2019rvt, Marks:2025jpt}.

Boson stars in anti-de~Sitter (AdS) spacetime have been studied for at least the past two decades~\cite{Astefanesei:2003qy, Radu:2012yx, Buchel:2013uba, Maliborski:2013ula, Brihaye:2014bqa, Fodor:2015eia, Chrusciel:2017uor, Liu:2020uaz, Guo:2020bqz, Herdeiro:2024myz, Liu:2025fcs, Zhao:2025yhy}.  
One motivation for considering these systems is that they allow the study of strong gravity scenarios in a  distinct  setting, as AdS naturally provides confinement through its ``reflective boundary''~\cite{hawking1975large, Bosch:2016vcp, Basu:2016srp, Peng:2017squ, Peng:2017gss, Ferreira:2017tnc}.
In fact, this confinement allows the existence of boson star solutions for massless scalar fields, which is not possible without the negative cosmological constant.
Perhaps more importantly, boson stars and other scalar field distributions in AdS have been considered in the context of conformal field theory (CFT), motivated by the AdS/CFT correspondence~\cite{Witten:1998qj, Maldacena:1997re}, mostly to ascertain their stability~\cite{Bizon:2011gg, PhysRevD.84.085021, Buchel:2012uh, Buchel:2013uba}.
Although ``total'' instability has been reported in~\cite{Bizon:2011gg, PhysRevD.84.085021}, it was soon after found that boson stars in AdS can actually be stable, at least in some regions of the parameter space~\cite{Dias:2012tq, Buchel:2013uba}.

In previous work~\cite{Alcubierre:2018ahf}, we presented a generalization of the standard boson stars, called $\ell$-boson stars, which consists of configurations of $2\ell+1$ scalar fields with angular momentum number $\ell$, all with the same amplitude, such that the stress energy-momentum tensor, and hence also the spacetime, is spherically symmetric, even though the individual scalar fields are not.  
Despite the seemingly arbitrary scalar field combinations that give rise to $\ell$-boson stars, we have shown that they arise naturally in a semiclassical context~\cite{Alcubierre:2022rgp}. 
Many properties of $\ell$-boson stars have been studied~\cite{Alcubierre:2018ahf, Alcubierre:2021psa}, including their linear~\cite{Alcubierre:2021mvs} and non-linear~\cite{Alcubierre:2019qnh} stability in spherical symmetry, and numerical 3D evolutions have also been performed~\cite{Jaramillo:2020rsv, Jaramillo:2022zwg}. Stability analyses have also been carried out in the non-relativistic limit~\mbox{\cite{Roque:2023sjl, Nambo:2023yut}}.

Regarding the stability of $\ell$-boson stars, we have seen a behavior similar to that in the standard, $\ell=0$, cases.
In the parameter space region starting from boson stars in the test field limit, and all the way until a mass maximum is reached (which we will call the {\em first region}), all ground state solutions are stable.
Beyond this point (which we will call the {\em second region}), the solutions become unstable.  
This is also a common behavior seen for other types of stars.
To be more precise, various alternating regions of stable and unstable solutions are sometimes present, delimited by the local maxima.
Such is the case even with some models that are unrelated to boson stars, like those describing white dwarfs and neutron stars~\cite{wald_general_1984}.

An interesting property to study is the existence of solutions containing zones with unstable circular orbits, with no circular orbits, and with light rings.
Typically, light rings appear in pairs, a stable innermost one and an unstable outermost one.
It was reported in~\cite{Cunha:2017qtt, Guo:2021bcw} that such light rings can appear only for boson stars that are themselves unstable.
That is, the light ring pairs appear, if at all, only in the second region, past the maximum mass.
However, according to recent work~\cite{Marks:2025jpt}, which considers solitonic boson stars in the thin-shell regime, stable boson stars of some particular types may actually admit light ring pairs.

In this article, we present $\ell$-boson stars in AdS: new solutions of the Einstein-Klein-Gordon equations with a negative cosmological constant ($\Lambda<0$) in spherical symmetry, describing boson stars with angular momentum number~$\ell$. 
As with the standard boson stars, these solutions consist of a scalar field with harmonic time-dependency and a static spacetime.
Asymptotically, as $r\to \infty$, the scalar field vanishes and the Schwarzschild-AdS spacetime is recovered, giving rise to asymptotically AdS solutions.  
We study the properties of these new solutions, emphasizing the differences with their counterparts with $\ell=0$ and with those having $\Lambda=0$. 
In particular, we describe the solutions' spectrum, compactness, and geodesic motion, including the existence of light rings, which, as we show here,  can exist even in the first region for large enough $\ell$.

The remainder of this article is organized as follows. After presenting the $\ell$-boson star model in Sec.~\ref{model} (and Appendix~\ref{compact}), solutions are obtained in Sec.~\ref{results} by solving numerically the resulting system of ordinary differential equations, where we also study some of their properties.
We conclude with a summary and discussion in Sec.~\ref{Sec:Conclusions}.

Throughout the article, we use the metric signature $(-,+,+,+)$ and units where the speed of light and the reduced Planck constant are $c=\hbar=1$.
On the other hand, we keep the gravitational constant $G$ explicit in most of the article, setting  it to $G=1$ only for the numerical results presented in Sec.~\ref{results}.

\section{Model} \label{model}
In this section, we present our model, derive the equations and asymptotic behavior, and define some quantities that will be useful when describing the solutions. 

We consider an odd number of complex non-interacting scalar fields $\Phi_{m}$, all of them with the same  mass, $\mu$, on a  (3+1) dimensional spacetime with negative cosmological constant, $\Lambda<0$.
The action can be written as
\begin{equation}
\begin{split}
  S &=  \int d^4x \sqrt{-g} \left[\frac{1}{16\pi G} \left(R - 2\Lambda\right) + \mathcal{L}_\Phi \right] ,\\
  \mathcal{L}_\Phi &= - \frac{1}{2} \sum_m \left( \nabla_\mu \Phi_m \nabla^\mu \Phi^*_m + \mu^2 \left| \Phi_m \right|^2 \right),
\end{split}
\end{equation}
where $R$ is the Ricci scalar and a star (${}^*$) denotes complex conjugate.

One can then obtain the Einstein-Klein-Gordon equations, 
\begin{equation}
  \begin{split}
    R_{\mu \nu} - \frac{1}{2} g_{\mu \nu} R + \Lambda\, g_{\mu \nu} &= 8 \pi G\, T_{\mu \nu}, \\
    \nabla_\mu \nabla^\mu \Phi_m - \mu^2  \Phi_m &= 0,
  \end{split}
\end{equation}
where the stress energy-momentum tensor is given by
\begin{eqnarray}
  T_{\mu\nu}&=&\frac{1}{2}\sum_m \left[ \nabla_\mu\Phi_m^*\nabla_\nu\Phi_m + \nabla_\mu\Phi_m\nabla_\nu\Phi_m^* \right. 
  \left.  - g_{\mu\nu} \left( \nabla_\alpha\Phi_m^*\nabla^\alpha\Phi_m + \mu^2 \Phi_m^*\Phi_m \right) \right] .
\label{Tmunu}
\end{eqnarray}

We will look for solutions of the form 
\begin{equation}   \label{ansatz}
  \Phi_m = \Phi_{\ell m}(t,r,\theta,\varphi) := \psi_\ell(r)\, e^{i\omega_\ell t}\, Y^{\ell m}(\theta,\varphi)
\end{equation}
for a given positive integer $\ell$ and with $m=-\ell,...,\ell$ labeling the $2\ell+1$ fields. 
Here, $Y^{\ell m}(\theta,\varphi)$ are the standard spherical harmonics.
Note that we choose the radial and temporal part to be the same for all the $2\ell+1$ fields.
This choice allows us to search for spherically symmetric spacetimes, in which the stress energy-momentum tensor~\eqref{Tmunu} is spherically symmetric even though the individual field components are not spherically symmetric themselves.
Regarding the time dependence in the ansatz~\eqref{ansatz}, we note that, as it is also well known to be the case in asymptotic flatness, no static boson stars can exist in the case with a negative cosmological constant~\cite{Peng:2019uzw}. Nevertheless, this simple time dependence in the scalar field still leads to static spacetime solutions.

It will be useful to follow two separate approaches: The first one is the same as in our previous work~\cite{Alcubierre:2018ahf}, while the second one is based on~\cite{Buchel:2013uba, PhysRevD.84.085021}, where, besides defining the metric functions differently, the authors choose a compact spatial coordinate.  
Next, we will derive the equations and asymptotic behavior using the first approach, and leave the second one for Appendix~\ref{compact}.
The reason for using both the non-compact and the compact formulations is twofold:
First, it helps when comparing results with different previous works that use one formulation or the other.
The second reason is that the shooting algorithm used (see Sec.~\ref{results}) fails to find solutions in certain regions of the parameter space, but sometimes it fails in different  regions for each formulation.
Hence, using both formulations allows one to explore larger regions of the parameter space.

Assuming a static, spherically symmetric spacetime,  we  choose a metric of the form
\begin{equation}
  ds^2 = -\alpha(r)^2 dt^2 + a(r)^2 dr^2 + r^2 \left(  d\theta^2 + \sin^2(\theta)\, d\varphi^2 \right) 
\end{equation}
for some metric functions $\alpha(r)$ and $a(r)$. 

The non-null components of the stress energy-momentum tensor then take the form
\begin{equation} \label{Tmunu2} 
  \begin{split} 
    {T^t}_t &= - \frac{(2\ell+1)}{8\pi} \left[ \frac{{\psi_\ell^\prime}^2}{a^2} + \frac{\omega_\ell^2}{\alpha^2}\psi_\ell^2 + \left( \mu^2 + \frac{\ell(\ell+1)}{r^2}  \right) \psi_\ell^2 \right] ,  \\
    {T^r}_r &=   \frac{(2\ell+1)}{8\pi} \left[ \frac{{\psi_\ell^\prime}^2}{a^2} + \frac{\omega_\ell^2}{\alpha^2}\psi_\ell^2- \left( \mu^2 + \frac{\ell(\ell+1)}{r^2}  \right) \psi_\ell^2 \right] ,  \\
    {T^\theta}_\theta =  {T^\varphi}_\varphi &=  \frac{(2\ell+1)}{8\pi} \left[ - \frac{{\psi_\ell^\prime}^2}{a^2} + \frac{\omega_\ell^2}{\alpha^2}\psi_\ell^2-  \mu^2  \psi_\ell^2 \right] , 
  \end{split}
\end{equation}
where primes (${}^\prime$) denote derivative with respect to $r$. While deriving Eqns.~\eqref{Tmunu2} we have used the addition theorem of the spherical harmonics to make explicit the angular independence. 

The Einstein-Klein-Gordon equations can be written as
\begin{subequations}
  \label{KG}
  \begin{align}
    a^\prime &= \frac{a}{2}      \bigg\{ -\frac{a^2-1}{r} + G_\ell\, r \bigg[ \left(  \mu^2 + \frac{\omega_\ell^2}{\alpha^2} + \frac{\ell(\ell+1)}{r^2} \right) a^2 \psi_\ell^2 \bigg. \bigg. 
               \Big. \Big.  + {\psi_\ell^\prime}^2 \Big] + \Lambda\, r\, a^2  \Big\}, \\
    \alpha^\prime &= \frac{\alpha}{2} \bigg\{  \frac{a^2-1}{r} + G_\ell\, r \bigg[ \left( -\mu^2 + \frac{\omega_\ell^2}{\alpha^2} - \frac{\ell(\ell+1)}{r^2} \right) a^2 \psi_\ell^2 \bigg. \bigg. 
      \bigg. \bigg.  + {\psi_\ell^\prime}^2 \bigg] - \Lambda\, r\, a^2  \bigg\} , \\
    \label{eqpsi}
    \psi_\ell^{\prime\prime} &= a^2 \bigg[ G_\ell\, r \left( \mu^2 + \frac{\ell(\ell+1)}{r^2} \right) \psi_\ell^2 \psi_\ell^\prime \bigg. 
     \bigg. + \left( \mu^2-\frac{\omega_\ell^2}{\alpha^2} + \frac{\ell(\ell+1)}{r^2} \right) \psi_\ell + \Lambda\, r\, \psi_\ell^\prime \bigg] - \frac{a^2+1}{r} \psi_\ell^\prime , 
  \end{align}
\end{subequations}
where we have defined the constant $G_\ell := (2\ell+1)\,G$.

Given the behavior of the field $\psi_\ell(r)$ for $r\rightarrow 0$, which goes to zero as $r^\ell$, it will be convenient to impose a change of variable that  facilitates setting boundary conditions at $r=0$ (see~\cite{Alcubierre:2018ahf} and~Sec.~\ref{BC}). We define $u_\ell(r)$ such that
\begin{equation} \label{defu}
  \psi_\ell(r) = r^\ell u_\ell(r)
\end{equation}
and  rewrite Eqns.~\eqref{KG} in terms of $u_\ell(r)$ to perform the actual numerical integrations as shown in Sec.~\ref{results}. 
One can see that these equations reduce to the expected ones when either $\Lambda=0$~\cite{Alcubierre:2018ahf} or $\mu=\ell=0$~\cite{Buchel:2013uba}.\footnote{After transforming to the different variables used in each reference. Also note that the definition of the scalar field in~\cite{Buchel:2013uba} differs by a factor $\sqrt{4\pi G}$. One can account for that discrepancy by simply replacing $4\pi G=1$ in Eqns.~\eqref{KG}.
}

We now define the mass function as
\begin{equation}
  \mathcal{M}(r) =  4\pi\, \int_0^r \tilde{r}^2 \, \rho(\tilde{r})\, d\tilde{r} , 
\end{equation}
where we have also defined the density 
$\rho(r) = - G\, {T^t}_t.$
We then have
\begin{equation}
  \lim_{r\rightarrow\infty} \mathcal{M}(r) = M,
\end{equation}
where $M$, defined below in Eqn.~\eqref{metricinf}, can be interpreted as the boson star mass~\cite{Buchel:2013uba}.

We define the maximum compactness $C_{\rm m}$ in terms of the mass function as
\begin{equation}
  C_{\rm m} = \max_{r>0} \left\{ \frac{\mathcal{M}(r)}{r} \right\} = \frac{M_{\rm m}}{r_{\rm m}}\, ,
  \label{eq:Cm}
\end{equation}
where we have also defined the radius of maximum compactness $r_{\rm m}$ as the location of the maximum, and the mass of maximum compactness $M_{\rm m}=\mathcal M(r_{\rm m})$.
Other commonly used definitions are $r_{99}$ and $C_{99}$, where $r_{99}$ is the radius containing $99\%$ of the total mass, and the corresponding compactness is
$C_{99}:=M/r_{99}$.
Both sets of definitions are meaningful and, in general, describe different aspects of the stars. They tend to differ more for stars with a dense ``core'' and a slowly decaying ``tail'', in which case the first set is more descriptive of the core, while the second set is more descriptive of the  whole star.

We now define the AdS radius $L>0$  as
$L:=\sqrt{-3/\Lambda}$,
and note that the solutions have the simple rescaling properties
\begin{equation}
  \begin{split}
  (L, r) &\mapsto \lambda (L, r)  \\
  (\mu, \omega_\ell) &\mapsto \lambda^{-1} (\mu, \omega_\ell) \, ,\\
    (a,\alpha, \psi_\ell) & \mapsto  (a,\alpha, \psi_\ell) 
  \end{split}
\end{equation}
which allow one to express most results in an $L$-independent form.

\subsection{Asymptotic behavior and boundary conditions}
\label{BC}

We want to obtain solutions that are regular at $r=0$ and with a rapidly decaying scalar field for $r\rightarrow\infty$.  
Starting with the left boundary case, we simply impose regularity at $r=0$, obtaining
\begin{equation} \label{eq:BC}
  a(0) = 1 ,\, 
  \alpha(0) = \alpha_0 ,\, 
  u(0) = u_0 ,\, 
  u^\prime (0) = 0 ,
\end{equation}
where $\alpha_0$ and $u_0$ are constants.

Besides the conditions at exactly $r=0$, we will also need to know the behavior of some functions for $r\gtrsim 0$ to avoid dividing by $0$ when evaluating the right-hand side of Eqns.~\eqref{KG}.
Keeping only the first couple of nonzero terms, we have
\begin{subequations}
  \begin{align}
    a(r) &\approx 
    \begin{dcases}
      1 + \left[  \frac{\Lambda}{6}  + \left( \mu^2 + \frac{\omega_\ell^2}{\alpha_0^2} \right) \frac{G_\ell u_0^2}{6}   \right]  r^2    & \ell = 0 \\
      1 + \left( \frac{\Lambda}{6} +  \frac{G_\ell u_0^2}{2} \right) r^2    &\ell = 1  \\
      1 +  \frac{\Lambda}{6} \,r^2      & \ell\ge 2   
    \end{dcases} ,  \\
    \frac{\alpha(r)}{\alpha_0} &\approx
    \begin{dcases}
      1 -  \left[ \frac{\Lambda}{6} + \left(\mu^2-2\frac{\omega_\ell^2}{\alpha_0^2}\right) \frac{G_\ell u_0^2}{6}   \right] r^2  & \ell=0 \\
      1 - \frac{\Lambda}{6}  r^2   & \ell\ge1 
    \end{dcases},  \\
    \frac{u_\ell(r)}{u_0} &\approx   1 + 
    \frac{  \ell(\ell+3) \Lambda + 3 \left( \mu^2 - \frac{\omega_\ell^2}{\alpha_0^2}  \right) + 9 G_\ell \delta_{\ell,1}  u_0^2  }{6 (2\ell+3)} r^2  . \label{cl}
  \end{align} 
\end{subequations}
Hence, at $r=0$, we replace the right-hand side of the equations by the expressions
\begin{equation}
  a^\prime(0) = 0, \;  
  \alpha^\prime(0) = 0, \; 
  u^\prime(0) = 0, \;  
  u^{\prime\prime}(0) = 2 c_\ell u_0 ,
\end{equation}
where $c_\ell$ is the coefficient multiplying $r^2$ in Eqn.~\eqref{cl}, avoiding in this way division by zero in the numerical code.

Moving on now to the right boundary,
for $r\rightarrow \infty$ we look for solutions with rapidly decaying $\psi_\ell(r)$ and assume the metric takes the form 
\begin{eqnarray}
  \label{metricinf}
  ds^2 &=& -\left( 1-\frac{2GM}{r}-\frac{\Lambda}{3}r^2 \right) \alpha_1^2 dt^2 
  + {\left( 1-\frac{2GM}{r}-\frac{\Lambda}{3}r^2 \right)}^{-1} dr^2 + r^2 d\Omega^2,
\end{eqnarray}
so that the metric functions are
\begin{equation}
  \label{alphainf}
  \frac{\alpha(r)}{\alpha_1} = \frac{1}{a(r)} = {\left( 1-\frac{2GM}{r}-\frac{\Lambda}{3}r^2 \right)}^\frac{1}{2},
\end{equation}
where $M$ and $\alpha_1$ are constants.

Keeping only the dominating terms in Eqn.~\eqref{alphainf} we have, for $\Lambda\neq0$,
\begin{equation}
\label{el}
  \frac{\alpha^2(r)}{\alpha_1^2} = \frac{1}{a^2(r)} =  -\frac{\Lambda}{3}r^2 .
\end{equation}
Substituting Eqn.~\eqref{el} into the equation for $\psi_\ell(r)$, Eqn.~\eqref{eqpsi}, and keeping only leading terms we obtain 
\begin{equation}
  \label{psiinf}
  \psi_\ell(r) \sim r^{-\left( \frac{3}{2}+\frac{1}{2} \sqrt{\frac{12\,\mu^2}{-\Lambda}+9} \right)}
\end{equation}
as the decaying solution for $\Lambda\neq0$. 
This coincides with~\cite{Astefanesei:2003qy}.
Note that the expression inside the parentheses is greater than or equal to $3$, being exactly $3$ only for $\mu=0$.
In that particular case, this is consistent with the first term in the expansion of~\cite{Buchel:2013uba}.

Similarly, for $\Lambda=0$, $\mu\neq 0$ we derive the expression
\begin{equation}
  \label{Lambda0}
  \psi_\ell (r) \sim \frac{ e^{ - \sqrt{\mu^2-\frac{\omega_\ell^2}{\alpha_1^2}} \, r } }{r}.
\end{equation}

The particular case where both $\Lambda=0$ and $\mu=0$ is of less interest for us since it does not admit boson star solutions, and we will simply disregard it.

Finally, to impose boundary conditions at some $r/L\gg1$ in our numerical code, we actually need the value of $u_\ell^\prime$ expressed in terms of $u_\ell$, which is given by
\begin{equation} \label{eq:BCinf}
  \frac{u_\ell^\prime(r)}{u_\ell(r)} \approx 
  \begin{dcases}
   - \frac{\frac{3}{2}+\frac{1}{2} \sqrt{\frac{12\,\mu^2}{-\Lambda}+9} +\ell }{r}  , & \Lambda < 0 \\
      - \sqrt{\mu^2-\frac{\omega_\ell^2}{\alpha_1^2}} - \frac{\ell+1}{r}   , & \Lambda=0,\, \mu\neq 0 
  \end{dcases} ,
\end{equation}
where we set initially $\alpha_1 = a(r)\alpha(r)$, which is constant for $r/L\gg1$, and later rescale the solutions so that $\alpha_1=1$, as we explain next.

Due to the solutions' rescaling properties, one can arbitrarily fix either $\alpha_0$ or $\alpha_1$.
We will set $\alpha_0=1$ for the numerical integrations, and afterwards rescale the solutions so that $\alpha_1=1$.
In this way, one obtains the Schawarzschild-AdS metric in one of its standard forms at $r\rightarrow \infty$, that of Eqn.~\eqref{metricinf} with $\alpha_1=1$. 
The constant $u_0$ is then the last remaining free parameter after fixing $\mu$, $\ell$, and $\Lambda$. 
Finally, imposing these boundary conditions  will determine the discrete spectrum $\omega_{\ell, n}$.

\subsection{Low-mass spectrum}
It can be useful to obtain the spectrum, $\omega_{\ell,n}$, in the low mass limit. For instance, it can help in the choice of  an appropriate initial guess when using a shooting method to solve the equations (see Sec.~\ref{results}).
It can also be useful for checking the consistency of the numerical solutions in that limit.
We proceed by performing a linear expansion of the form
\begin{equation}
  \begin{split}
    a(r) &=  \left(1-\frac{\Lambda }{3} r^2\right)^{-\frac{1}{2}} + \lambda a^{(1)}(r) + \lambda^2 a^{(2)}(r) + \cdots  \\
    \alpha(r) &= \left(1-\frac{\Lambda }{3} r^2\right)^\frac{1}{2} + \lambda \alpha^{(1)}(r) + \lambda^2 \alpha^{(2)}(r) + \cdots  \\
    \psi_\ell(r) &= \lambda \psi_\ell^{(1)}(r) + \lambda^2 \psi_\ell^{(2)}(r) + \cdots  \\
    \omega_\ell &= \omega_\ell^{(0)} + \lambda \omega_\ell^{(1)} + \lambda^2 \omega_\ell^{(2)} + \cdots
  \end{split}
\end{equation}
Replacing these expansions into Eqns.~\eqref{KG} and solving for $\psi_\ell^{(1)}(r)$ one obtains analytic solutions in terms of hypergeometric functions. Imposing the proper boundary conditions restricts the possible frequency values to those in the expression
\begin{equation}
  \label{spectrum}
  L \, \omega_{\ell,n}^{(0)} = 2n + \ell + \frac{3}{2}  + \frac{1}{2} \sqrt{4 (L \mu)^2 + 9}  \, ,
\end{equation}
where $n=0$, $1$, $2$, $\cdots$.

We have verified that our numerical solutions indeed satisfy this expression in the low-mass limit. See, for instance, Fig.~\ref{f:Mw} for some examples with $n=0$.
Eqn.~\eqref{spectrum} is also consistent with the simple expression obtained in~\cite{Buchel:2013uba} for the case with $\ell=0$ and $\mu=0$.

\section{Solutions and properties}
\label{results}
  
The solutions are obtained by numerically integrating Eqns.~\eqref{KG} after substituting Eqn.\eqref{defu} and imposing boundary conditions as in Eqns.~\eqref{eq:BC}, \eqref{alphainf}, and~\eqref{eq:BCinf}.
Or, when using the  compact formulation, the solutions are obtained by integrating~\eqref{KGc} after substituting~\eqref{vdef} and imposing boundary conditions in Eqns.~\eqref{BCx1} and~\eqref{BCx2}.
To do that, we use a shooting algorithm based on the one described in~\cite{Press1992Numerical}, but replacing the solver for the adaptive step one presented in~\cite{Radhakrishnan1993}, which provides high accuracy at low computational cost.
The numerical code with this implementation was first presented and tested in~\cite{Megevand:2007uy}.
The code is run using double precision and a tolerance set to $10^{-14}$, which ensures integration with many significant figures of accuracy. The boundary conditions are exact, except for the right-hand one in the non-compact formulation. The effects of that boundary condition are controlled by setting it at progressively larger radii until no significant changes are seen in the solutions.
All quantities are hence presented with very high accuracy, except for the compactness, as explained next. The step adaptive code usually  utilizes a very large number of grid points, which would make it difficult to store the hundreds of solutions obtained. For that reason, only a smaller grid, of 20,000 equidistant points, is kept. The compactness is evaluated in terms of $r_m$ in Eqn.~{\eqref{eq:Cm}}, using the values in the smaller grid, resulting in some cases in errors of up to 4\%.
In this section, we set $G=1$. 
 
We will concentrate mainly on ``ground state'' solutions, that is, those with no nodes for $r>0$, and labeled with $n=0$.
``Excited state'' modes, for which $n>0$ counts the number of nodes, were shown to be unstable for other types of boson stars~\cite{Balakrishna:1997ej}, although some combinations of ground and excited states can again be stable~\cite{Bernal:2009zy}. In our case, a stability analysis is left for future work, so we cannot yet assert whether excited modes are unstable. Regardless, for simplicity, we will concentrate here mainly on studying the ground state solutions.

Exploiting the rescaling properties of the equations, we obtain and show the solutions in an $L$-independent form.
We study the cases with $n=0$, $\mu=0$ and $L\mu=1$, $\ell=0$ to $\ell=15$, and, for each of those, multiple values of $u_0$ such that all the first region and part of the second region are covered. 
Additionally, we found a few solutions with $n>0$, which are not shown in the figures presented here.

\begin{figure}[htb]
  \includegraphics[width=\figlen\textwidth]{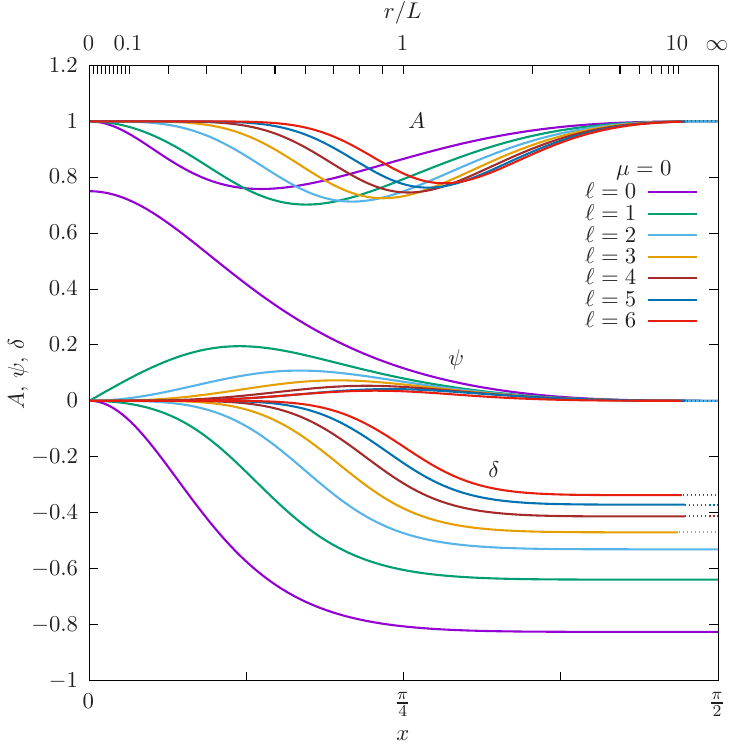}  
  \caption{Field profiles for $\mu=0$ in the maximum mass cases. 
    \label{f:fields}}
\end{figure}

\begin{figure}[htb]
  \includegraphics[width=\figlen\textwidth]{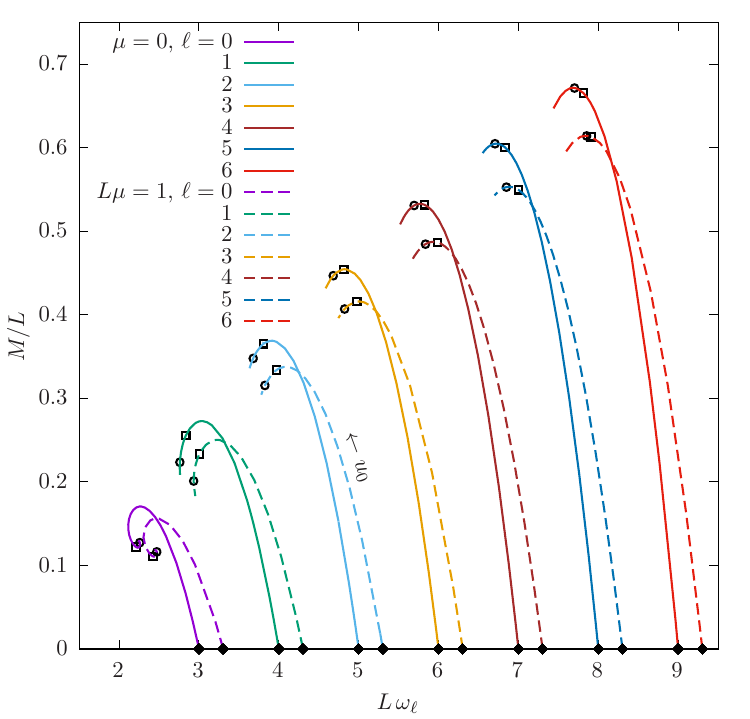}
  \caption{Total mass $M$ vs frequency $\omega_\ell$ for $\mu=0$ and $L \mu=1$.
    \label{f:Mw}}
\end{figure}

We start by showing some field profiles in Fig.~\ref{f:fields}.
We can see that some curves do not reach the figure's right boundary.
This is because those solutions were obtained numerically using the non-compact formulation, setting a boundary at finite, but large enough values of $r$.
However, those curves can be easily extended using the asymptotic expansions presented in Sec.~\ref{model} or in Sec.~\ref{compact}.
We show the extended curves with dotted lines. 
For clarity, we show here only some examples with $\mu=0$, $\ell=0$ to $\ell=6$, and $u_0$ such that each solution is that of maximum mass (the maxima seen in Fig.~\ref{f:Mw}).

\begin{figure}[htb]
  \includegraphics[width=\figlen\textwidth]{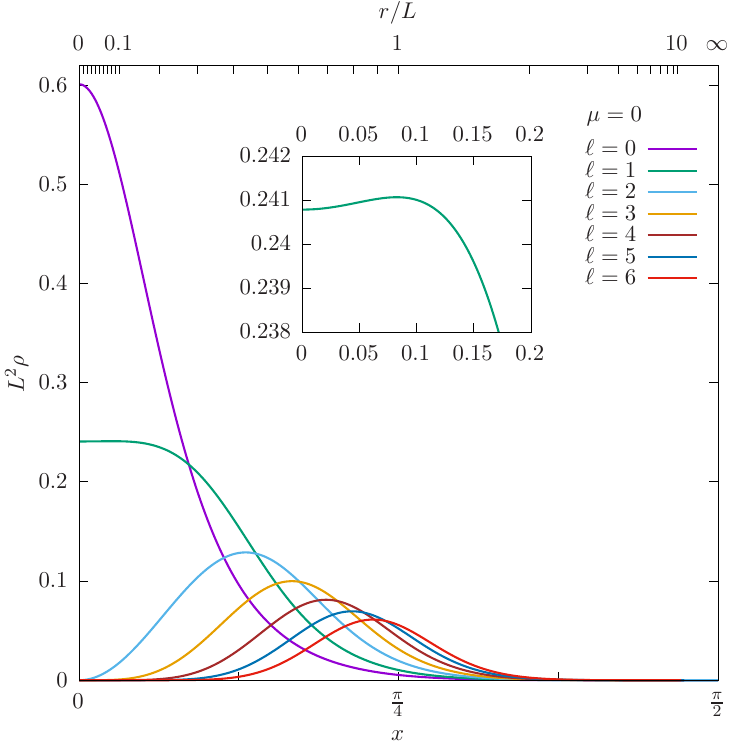}
  \caption{Density profiles for $\mu=0$ in the maximum mass cases. 
    \label{f:rho}} 
\end{figure}

Next, we show some density profiles in Fig.~\ref{f:rho}.
These correspond to the same solutions as those shown in the previous figure.
We see that the stars become wider for larger $\ell$, and display a thick-shell-like form when $\ell\ge 2$.
The almost hollow central region, with exactly zero density at $r=0$, also becomes larger with $\ell$.
On the other hand, the stars with $\ell=0$ and $\ell=1$ are qualitatively very different.
The case with $\ell=0$ is the only one with maximum density located at the center.
The case with $\ell=1$ also has a non-zero density at $r=0$, but its maximum is located at some $r>0$.
All the solutions have zero slope at the center.
  
In Fig.~\ref{f:Mw}, we show the values of mass $M$ vs. frequency $\omega_\ell$ of solutions obtained when varying the central value of the scalar field $u_0$ for  given values of $\mu$ and $\ell$.
The figure includes the cases with $\ell$ from $0$ to $6$, each with $\mu=0$ and with $L \mu=1$. Although solutions with larger values of $\ell$ have been found, we prefer to include in this figure, for clarity,  only solutions with the first few values.
The squares indicate the first solution (when starting with small $u_0$ and increasing it) for which there exist unstable circular orbits.
Similarly, the circles indicate the first solution with light rings.
We will discuss circular orbits in more detail later.
The low mass solutions correspond to small values of $u_0$.
In that limit, the values coincide with those given by Eqn.~\eqref{spectrum}, indicated in the figure with diamonds.
We have verified this is also true in cases with $n>0$, which are not shown in this figure.
The curves are qualitatively similar to those of the $\ell$-boson stars with $\Lambda=0$, with decreasing frequency and increasing mass up to a critical point as the scalar field amplitude increases.
The critical point corresponds in each case to the solution of maximum mass. This point usually indicates the transition from a stable region, to the right, to an unstable region, to the left.\footnote{Note that, even though boson star solutions in the first branch have been shown to be stable in other scenarios in which $\Lambda=0$, the AdS case is more subtle due to the existence of turbulent instabilities. However, as mentioned earlier, stable solutions still exist in some regions of the parameter space for solutions located in the first branch.}
Since we have not performed a stability analysis yet, we refer to these two regions as the {\em first} and {\em second} region, respectively. 
We also see that Eqn.~\eqref{spectrum} for the frequency is satisfied in the low-mass limit. This is true for all the cases we studied, including ones not shown in the figure, as those with $n>0$ and $\ell>6$.
Finally, we note that larger values of $\ell$ admit solutions with larger mass.

\begin{figure}[htb]
  \includegraphics[width=\figlen\textwidth]{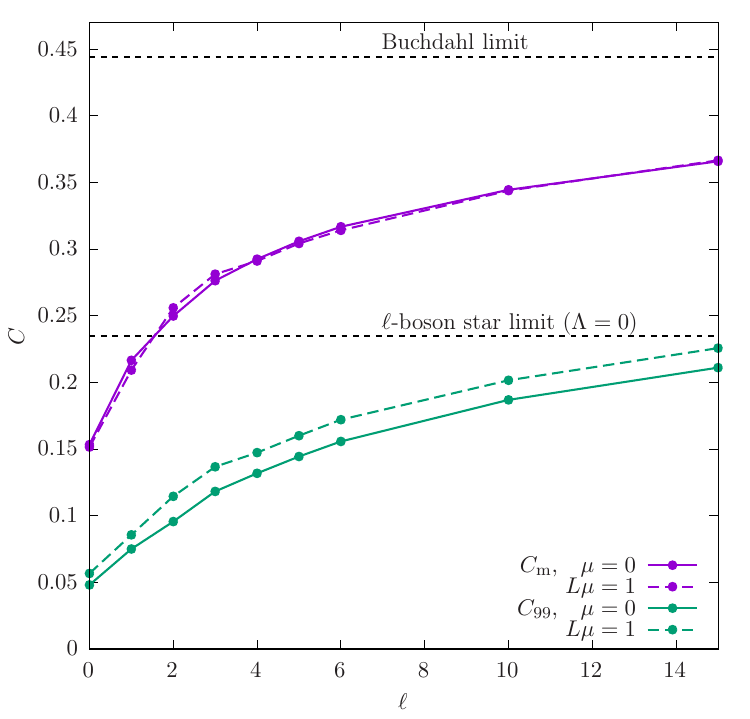}
  \caption{Compactness of maximum mass solutions vs. $\ell$.
    \label{f:Cml}}
\end{figure}

In order to compare solutions with different values of $\ell$, we decide to concentrate on the particular solutions with the maximum mass.
Fig.~\ref{f:Cml} shows how the compactness increases with $\ell$.
This fact might not be obvious since both the mass and radius increase, but they do so in such a way that the compactness always increases.
This is true for both definitions of compactness.
We see that $C_{\rm m}$ quickly reaches values that are much larger than its asymptotic limit for $\ell\rightarrow \infty$ found for $\ell$-boson stars in the case with $\Lambda=0$~\cite{Alcubierre:2018ahf}.
This limit is indicated in the figure with a black dashed line.
On the other hand, the compactness  remains below the Buchdahl limit of $4/9$~\cite{Buchdahl:1959zz}, at least for solutions with $\ell\le 15$, which is the maximum value we have studied for this work.

\begin{figure}[htb]
  \includegraphics[width=\figlen\textwidth]{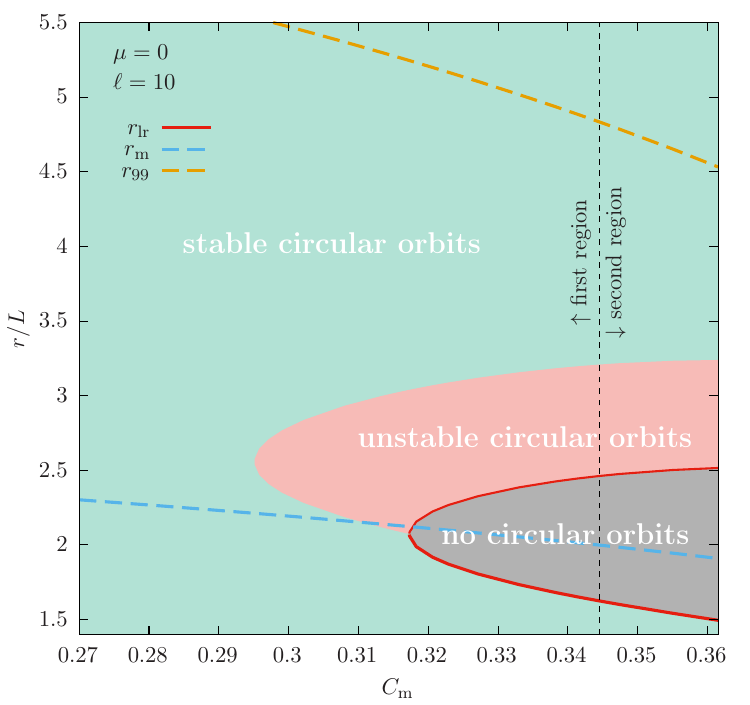}
  \caption{Types of circular orbits for solutions with $\mu=0$ and $\ell=10$.
    \label{f:LR}}
\end{figure}

Following the procedures shown, for instance, in~\cite{wald_general_1984, Barranco:2021auj, Alcubierre:2021psa}, we study the existence and stability of time-like and null circular orbits.
We show some properties of circular orbits for the case with $\mu=0$ in Fig.~\ref{f:LR}.
Each vertical line represents a solution with a given compactness $C_{\rm m}$.
The colored regions distinguish different types of circular orbits at the given radii: light green for stable orbits, light red for unstable ones, and gray where no circular orbits exist.
The two red lines indicate the location of the light rings, an unstable outermost one and a stable innermost one.
Note that, contrary to the case with $\Lambda=0$, solutions with light ring pairs exist in the first region.
For $\ell\ge4$ (not shown in the figure), we find that unstable circular orbits can exist in the first region.
In the $\Lambda=0$ case, this is true only for $\ell\ge 9$~\cite{Alcubierre:2021psa}.
Additionally, solutions with zones of no circular orbits, which are delimited by a pair of light rings, exist for $\ell\ge 6$.
Note that no solutions with such properties were found in~\cite{Alcubierre:2021psa}, where solutions with values up to $\ell=1600$ were studied.
The results obtained for $L\mu=1$ in this regard are the same as those  described for the cases with $\mu=0$.
We mentioned that light rings exist in the first region for $\ell\ge 6$.
For $\ell=6$, only solutions in a small range (of $u_0$) with $M<M_{\rm max}$ have light rings.
This range is larger the larger the value of $\ell$.
As an illustrative case, we choose to show in this figure that with $\ell=10$, which displays clearly all the relevant regions.
However, the figures for all values of $\ell$ are qualitatively very similar, with the only difference being the relative location of the vertical dashed line, which divides the first and second region.


\FloatBarrier

\section{Summary and discussion}\label{Sec:Conclusions}

We have presented $\ell$-boson stars in AdS spacetime, static solutions of the Einstein-Klein-Gordon equations with angular momentum number $\ell$.
They are parameterized by the three continuous parameters --the scalar field mass $\mu$, the AdS radius $L$, and the central field $u_0$--, as well as the two discrete values of the  angular momentum number $\ell$ and mode number $n$ (set mostly to zero here). 
Note that solutions with $\mu=0$ exist, which is not the case when $\Lambda=0$.
These stars inherit some properties from ($\ell=0$) boson stars in AdS, some others from ($\Lambda=0$) boson stars, and also show some properties not seen in any of those two cases.

Morphologically, they are thick-shell-like structures with a low-density central region, except only when $\ell\le 1$.
Both the star radius and the central region, as well as the mass and compactness, can be larger for larger $\ell$.  
Note, for instance, that for $\ell$ as low as $\ell=2$ we already find solutions that are more compact than any other one with $\Lambda=0$, even those with $\ell\rightarrow \infty$.
 
The very high compactness that can be achieved, larger than that in the cases with $\ell=0$ or $\Lambda=0$, gives rise to an interesting feature not present in those cases.
Starting at $\ell=6$, we find solutions in the first region that contain a pair of light rings, one of them being stable.
The location of the light rings delimits a zone where no circular orbits can exist. 
The existence of light rings in the first region is particularly interesting, as the first region usually contains stable solutions in most cases of similar (and even not so similar) models as that studied in the present work~\cite{Liebling:2012fv, Alcubierre:2021mvs, Alcubierre:2019qnh, wald_general_1984}. 
This result would be in opposition with that found in~\cite{Cunha:2017qtt, Guo:2021bcw} for the case with $\Lambda=0$, where only unstable solutions admit such light ring pairs.
Boson stars (without light rings) have previously been considered as black hole mimickers~\mbox{\cite{Schunck:2003kk, Liebling:2012fv}}. The possible existence of stable solutions with light rings could make the similarity with black holes even higher. 
However, given the negative cosmological constant, this aspect may actually be of more importance in the context of the AdS/CFT correspondence. See, for instance,~\mbox{\cite{Hashimoto:2023buz, Riojas:2023pew}}, where the (external) light ring plays a role in the study of quasi-normal modes for a black hole in AdS.
  
Finally, conducting a complete stability analysis for the $\ell$-boson stars in AdS would indeed be very interesting.
However, that is beyond the scope of the present work, and it will be left for a future project.


\acknowledgments
We thank the anonymous referee for their useful suggestions.

\appendix

\section{Compact formulation}
\label{compact}

In this appendix, we derive the main equations and study the asymptotic behavior using a compact formulation.
We consider a metric of the form
\begin{eqnarray}
  ds^2 &=& \frac{L^2}{\cos^2(x)}\bigg[  -\frac{A(x)}{{L^2} e^{2\delta(x)}} \, dt^2 +  \frac{1}{A(x)} \, dx^2 \bigg. 
   \bigg. + \sin^2(x) \left( d\theta^2 + \sin^2(\theta) \, d\varphi^2 \right)  \bigg],  
\end{eqnarray}
where $x \in [0,\frac{\pi}{2}]$ can be expressed in terms of the radial coordinate $r$ as
$\tan(x)=r/L$.
and we keep the same definition for the scalar field as in the non-compact formulation. Its spatial component, expressed as a function of $x$, is then
$\phi_\ell(x) = \psi_\ell(r(x))$.
We restrict to $\Lambda<0$ (so that $L^2=-3/\Lambda>0$) in this formulation.

The nonzero components of the stress energy-momentum tensor are
\begin{equation}
  \begin{split}
    \bar{T}^t{}_t &= - \frac{2\ell+1}{8\pi L^2} \left[ \cos^2(x) A\, {\phi_\ell^\prime}^2 + \left( \frac{ \tilde{\omega}_\ell^2}{\tilde{\alpha}^2} + \tilde{\mu}^2 + \frac{\ell(\ell+1)}{\tan^2(x)}\right) \phi_\ell^2 \right]  ,  \\
    \bar{T}^x{}_x &= \frac{2\ell+1}{8\pi L^2} \left[  \cos^2(x) A\, {\phi_\ell^\prime}^2 + \left( \frac{ \tilde{\omega}_\ell^2}{\tilde{\alpha}^2} - \tilde{\mu}^2 - \frac{\ell(\ell+1)}{\tan^2(x)}\right) \phi_\ell^2   \right]  ,  \\
    \bar{T}^\theta{}_\theta = \bar{T}^\varphi{}_\varphi &=  \frac{2\ell+1}{8\pi L^2}  \left[ \cos^2(x) A\, {\phi_\ell^\prime}^2 + \left( \frac{ \tilde{\omega}_\ell^2}{\tilde{\alpha}^2} -\tilde{\mu}^2\right) \phi_\ell^2\right]  ,
  \end{split}
\end{equation}
where  primes (${}^\prime$) now denote derivative with respect to $x$, and we have defined
\begin{equation}
  \tilde{\alpha}^2(x) := \frac{A(x)}{\cos^2(x) e^{2\delta(x)}},  \;
  \tilde{\mu} := L \mu, \;
  \tilde{\omega}_\ell := L \omega_\ell .
\end{equation}

The Einstein-Klein-Gordon equations are
\begin{equation}
  \label{KGc}
  \begin{split}
    A^\prime &= -G_\ell \bigg[ \left( \frac{ \tilde{\omega}_\ell^2}{\tilde{\alpha}^2} +\tilde{\mu}^2  + \frac{\ell(\ell+1)}{\tan^2(x)}   \right) \tan(x) \phi_\ell^2 \bigg. 
     \bigg. + \cos(x) \sin(x) A {\phi_\ell^\prime}^2 \bigg] 
    + \frac{\left(2\cos^2(x)-3\right)\left(A-1\right)}{\cos(x)\sin(x)} ,  \\
    \delta^\prime &= -G_\ell \tan(x) \left( \cos^2(x)\,{\phi_\ell^\prime}^2 + \frac{ \tilde{\omega}_\ell^2 }{\tilde{\alpha}^2} \,\frac{\phi_\ell^2}{A} \right) , \\
    \phi_\ell^{\prime\prime} &= \left( \delta^\prime - \frac{A^\prime}{A} - \frac{2}{\cos(x)\sin(x)} \right)\phi_\ell^\prime 
    + \left( -\frac{ \tilde{\omega}_\ell^2}{\tilde{\alpha}^2} +\tilde{\mu}^2  + \frac{\ell(\ell+1)}{\tan^2(x) } \right) \frac{\phi_\ell}{\cos^2(x) A} .    
  \end{split}
\end{equation}
As expected, these equations reduce to those in~\cite{Buchel:2013uba} when $\ell=\mu=0$. 

As in the non-compact formulation, it will be convenient to replace the variable $\phi_\ell(x)$ with one that makes explicit the behavior at the origin. We then define the function $v_\ell(x)$ as
\begin{equation}
  \label{vdef}
  \phi_\ell(x) = \tan^\ell(x) \, v_\ell(x).
\end{equation} 
Note that with this choice, we have 
$v_\ell(x) = L^\ell \, u_\ell(r)$.
For the numerical integration, we will write the equations in terms of $v_\ell(x)$. That is, replace Eqn.~\eqref{vdef} into Eqns.~\eqref{KGc}.

The equations in both formulations are obtained independently of each other.
However, we find that by transforming the equations in terms of \{$A(x)$, $\delta(x)$, $v(x)$\}  to equations in terms of \{$a(r)$, $\alpha(r)$, $u(r)$\} one obtains the same equations as in the non-compact formulation, as expected.

We define the mass function as
\begin{equation}
  \mathcal{M}(x) =  4\pi\, \int_0^x \frac{L^3\, \tan^2(\tilde{x})}{\cos^2(\tilde{x})} \, \bar{\rho}(\tilde{x})\, d\tilde{x} \, ,
\end{equation}
where
$\bar{\rho}(x) = - G\, \bar{T}^t{}_t$.
We then have
$\mathcal{M}\left(\frac{\pi}{2}\right) = M$,   
where $M$, defined in the asymptotic expansion of $A(x)$, Eqn.~\eqref{Ainf}, can be interpreted as the boson star mass.

Finally, to connect with the non-compact formulation, we present the relations between the metric functions:
\begin{equation}
  \begin{split}
    A(x) &= \left\{\left[1+\left(\frac{r}{L}\right)^2\right] a(r)^2\right\}^{-1} , \\
    \delta(x) &= - \ln \left[ \alpha(r) a(r) \right] .
  \end{split}
\end{equation}

We move on now to study the fields' behavior at the boundaries.
In the limit $x\rightarrow 0$ we obtain

\begin{equation}
  \begin{split}
    A(x) &\approx 
    \begin{dcases}
      1 - \frac{G_\ell}{3} \left(\tilde{\mu}^2+e^{2\delta_0}\tilde{\omega}_\ell^2\right) v_0^2  \, x^2  & \ell = 0 \\
      1 - G_\ell\, v_0^2  \, x^2  & \ell = 1  \\
      1   \, ,   & \ell\ge 2 
    \end{dcases} , \\
    \delta(x) &\approx
    \begin{dcases}
      \delta_0 - \frac{G_\ell}{2}\,e^{2\delta_0}\,\tilde{\omega}_\ell^2\, v_0^2  \, x^2  & \ell=0 \\
      \delta_0 -\frac{G_\ell}{2}   \, v_0^2   \, x^2   & \ell=1 \\
      \delta_0  & \ell\ge 2
    \end{dcases}, \\
    \frac{v_\ell(x)}{v_0} &\approx   1 +  
         \frac{ -\ell(\ell+3) + \tilde{\mu}^2-e^{2\delta_0}\tilde{\omega}_\ell^2 + 3 G_\ell v_0^2   \delta_{\ell,1} }{2(2\ell+3)} x^2  , 
  \end{split}
\end{equation}
where $\delta_{\ell,1}$ is the Kronecker delta.

These expansions coincide with~\cite{Buchel:2013uba} for the particular case shown there, in which $\ell=\mu=0$ (and $4\pi G=L=1$). 
The free parameters $\delta_0$ and $v_0$ can be written in terms of previously defined parameters in the other formulation as
$v_0 = L^\ell\, u_0$ and $\delta_0 = -\ln\left({\alpha_0}\right)$.

The boundary conditions at $x=0$ are
\begin{equation}
  \label{BCx1}
  A(0)=1, \; \delta(0)=\delta_0,\; v_\ell(0)=v_0 ,\; v^\prime(0)=0.
\end{equation}

In the limit $x\rightarrow \pi/2$ we obtain

\begin{equation} 
  \begin{split}  
    A(x) &\approx 1 - \frac{2 G M}{L} {\left(\frac{\pi}{2}-x\right)}^3 \, , \\
    \delta(x) &\approx \delta_1 =  -\ln\left({\alpha_1}\right) \, , \\
    \phi(x) &\sim {\left(\frac{\pi}{2}-x\right)}^{\frac{3}{2}+\frac{1}{2}\sqrt{4\tilde{\mu}^2+9}} \, .
  \end{split}
  \label{Ainf}
\end{equation}
Note that the exponent in
the last equation
is greater than or equal to $3$, being exactly $3$ only when $\mu=0$.
We choose the following boundary conditions at $x=\pi/2$:
\begin{equation}
  \label{BCx2}
  A\left(\frac{\pi}{2}\right)=1, \; \delta\left(\frac{\pi}{2}\right)=\delta_1, \; v\left(\frac{\pi}{2}\right) = 0. 
\end{equation}

Similar to the choice of $\alpha_0$ and $\alpha_1$ in the non-compact formulation, either $\delta_0$ or $\delta_1$ can be chosen arbitrarily. In the numerical integration, we initially fix $\delta_0$ and later rescale the solutions so that $\delta_1$ is such that $\alpha_1=1$. In this way, a solution with $v_0=L^\ell u_0$ is the same one as that with $u_0$ in the non-compact formulation.

 

\begin{thebibliography}{77}%
\makeatletter
\providecommand \@ifxundefined [1]{%
 \@ifx{#1\undefined}
}%
\providecommand \@ifnum [1]{%
 \ifnum #1\expandafter \@firstoftwo
 \else \expandafter \@secondoftwo
 \fi
}%
\providecommand \@ifx [1]{%
 \ifx #1\expandafter \@firstoftwo
 \else \expandafter \@secondoftwo
 \fi
}%
\providecommand \natexlab [1]{#1}%
\providecommand \enquote  [1]{``#1''}%
\providecommand \bibnamefont  [1]{#1}%
\providecommand \bibfnamefont [1]{#1}%
\providecommand \citenamefont [1]{#1}%
\providecommand \href@noop [0]{\@secondoftwo}%
\providecommand \href [0]{\begingroup \@sanitize@url \@href}%
\providecommand \@href[1]{\@@startlink{#1}\@@href}%
\providecommand \@@href[1]{\endgroup#1\@@endlink}%
\providecommand \@sanitize@url [0]{\catcode `\\12\catcode `\$12\catcode
  `\&12\catcode `\#12\catcode `\^12\catcode `\_12\catcode `\%12\relax}%
\providecommand \@@startlink[1]{}%
\providecommand \@@endlink[0]{}%
\providecommand \url  [0]{\begingroup\@sanitize@url \@url }%
\providecommand \@url [1]{\endgroup\@href {#1}{\urlprefix }}%
\providecommand \urlprefix  [0]{URL }%
\providecommand \Eprint [0]{\href }%
\providecommand \doibase [0]{https://doi.org/}%
\providecommand \selectlanguage [0]{\@gobble}%
\providecommand \bibinfo  [0]{\@secondoftwo}%
\providecommand \bibfield  [0]{\@secondoftwo}%
\providecommand \translation [1]{[#1]}%
\providecommand \BibitemOpen [0]{}%
\providecommand \bibitemStop [0]{}%
\providecommand \bibitemNoStop [0]{.\EOS\space}%
\providecommand \EOS [0]{\spacefactor3000\relax}%
\providecommand \BibitemShut  [1]{\csname bibitem#1\endcsname}%
\let\auto@bib@innerbib\@empty
\bibitem [{\citenamefont {Kaup}(1968)}]{Kaup:1968zz}%
  \BibitemOpen
  \bibfield  {author} {\bibinfo {author} {\bibfnamefont {D.~J.}\ \bibnamefont
  {Kaup}},\ }\bibfield  {title} {\bibinfo {title} {{Klein-Gordon Geon}},\
  }\href {https://doi.org/10.1103/PhysRev.172.1331} {\bibfield  {journal}
  {\bibinfo  {journal} {Phys. Rev.}\ }\textbf {\bibinfo {volume} {172}},\
  \bibinfo {pages} {1331} (\bibinfo {year} {1968})}\BibitemShut {NoStop}%
\bibitem [{\citenamefont {Ruffini}\ and\ \citenamefont
  {Bonazzola}(1969)}]{Ruffini:1969qy}%
  \BibitemOpen
  \bibfield  {author} {\bibinfo {author} {\bibfnamefont {R.}~\bibnamefont
  {Ruffini}}\ and\ \bibinfo {author} {\bibfnamefont {S.}~\bibnamefont
  {Bonazzola}},\ }\bibfield  {title} {\bibinfo {title} {{Systems of
  selfgravitating particles in general relativity and the concept of an
  equation of state}},\ }\href {https://doi.org/10.1103/PhysRev.187.1767}
  {\bibfield  {journal} {\bibinfo  {journal} {Phys. Rev.}\ }\textbf {\bibinfo
  {volume} {187}},\ \bibinfo {pages} {1767} (\bibinfo {year}
  {1969})}\BibitemShut {NoStop}%
\bibitem [{\citenamefont {Colpi}\ \emph {et~al.}(1986)\citenamefont {Colpi},
  \citenamefont {Shapiro},\ and\ \citenamefont {Wasserman}}]{Colpi:1986ye}%
  \BibitemOpen
  \bibfield  {author} {\bibinfo {author} {\bibfnamefont {M.}~\bibnamefont
  {Colpi}}, \bibinfo {author} {\bibfnamefont {S.~L.}\ \bibnamefont {Shapiro}},\
  and\ \bibinfo {author} {\bibfnamefont {I.}~\bibnamefont {Wasserman}},\
  }\bibfield  {title} {\bibinfo {title} {{Boson Stars: Gravitational Equilibria
  of Selfinteracting Scalar Fields}},\ }\href
  {https://doi.org/10.1103/PhysRevLett.57.2485} {\bibfield  {journal} {\bibinfo
   {journal} {Phys. Rev. Lett.}\ }\textbf {\bibinfo {volume} {57}},\ \bibinfo
  {pages} {2485} (\bibinfo {year} {1986})}\BibitemShut {NoStop}%
\bibitem [{\citenamefont {Friedberg}\ \emph {et~al.}(1987)\citenamefont
  {Friedberg}, \citenamefont {Lee},\ and\ \citenamefont
  {Pang}}]{Friedberg:1986tp}%
  \BibitemOpen
  \bibfield  {author} {\bibinfo {author} {\bibfnamefont {R.}~\bibnamefont
  {Friedberg}}, \bibinfo {author} {\bibfnamefont {T.~D.}\ \bibnamefont {Lee}},\
  and\ \bibinfo {author} {\bibfnamefont {Y.}~\bibnamefont {Pang}},\ }\bibfield
  {title} {\bibinfo {title} {{MINI - SOLITON STARS}},\ }\href
  {https://doi.org/10.1103/PhysRevD.35.3640} {\bibfield  {journal} {\bibinfo
  {journal} {Phys. Rev. D}\ }\textbf {\bibinfo {volume} {35}},\ \bibinfo
  {pages} {3640} (\bibinfo {year} {1987})}\BibitemShut {NoStop}%
\bibitem [{\citenamefont {Gleiser}(1988)}]{Gleiser:1988rq}%
  \BibitemOpen
  \bibfield  {author} {\bibinfo {author} {\bibfnamefont {M.}~\bibnamefont
  {Gleiser}},\ }\bibfield  {title} {\bibinfo {title} {{Stability of Boson
  Stars}},\ }\href {https://doi.org/10.1103/PhysRevD.38.2376} {\bibfield
  {journal} {\bibinfo  {journal} {Phys. Rev. D}\ }\textbf {\bibinfo {volume}
  {38}},\ \bibinfo {pages} {2376} (\bibinfo {year} {1988})},\ \bibinfo {note}
  {[Erratum: Phys.Rev.D 39, 1257 (1989)]}\BibitemShut {NoStop}%
\bibitem [{\citenamefont {Lee}\ and\ \citenamefont {Pang}(1989)}]{Lee:1988av}%
  \BibitemOpen
  \bibfield  {author} {\bibinfo {author} {\bibfnamefont {T.~D.}\ \bibnamefont
  {Lee}}\ and\ \bibinfo {author} {\bibfnamefont {Y.}~\bibnamefont {Pang}},\
  }\bibfield  {title} {\bibinfo {title} {{Stability of Mini - Boson Stars}},\
  }\href {https://doi.org/10.1016/0550-3213(89)90365-9} {\bibfield  {journal}
  {\bibinfo  {journal} {Nucl. Phys. B}\ }\textbf {\bibinfo {volume} {315}},\
  \bibinfo {pages} {477} (\bibinfo {year} {1989})}\BibitemShut {NoStop}%
\bibitem [{\citenamefont {Seidel}\ and\ \citenamefont
  {Suen}(1990)}]{Seidel:1990jh}%
  \BibitemOpen
  \bibfield  {author} {\bibinfo {author} {\bibfnamefont {E.}~\bibnamefont
  {Seidel}}\ and\ \bibinfo {author} {\bibfnamefont {W.-M.}\ \bibnamefont
  {Suen}},\ }\bibfield  {title} {\bibinfo {title} {{Dynamical Evolution of
  Boson Stars. 1. Perturbing the Ground State}},\ }\href
  {https://doi.org/10.1103/PhysRevD.42.384} {\bibfield  {journal} {\bibinfo
  {journal} {Phys. Rev. D}\ }\textbf {\bibinfo {volume} {42}},\ \bibinfo
  {pages} {384} (\bibinfo {year} {1990})}\BibitemShut {NoStop}%
\bibitem [{\citenamefont {Liddle}\ and\ \citenamefont
  {Madsen}(1992)}]{Liddle:1992fmk}%
  \BibitemOpen
  \bibfield  {author} {\bibinfo {author} {\bibfnamefont {A.~R.}\ \bibnamefont
  {Liddle}}\ and\ \bibinfo {author} {\bibfnamefont {M.~S.}\ \bibnamefont
  {Madsen}},\ }\bibfield  {title} {\bibinfo {title} {{The Structure and
  formation of boson stars}},\ }\href
  {https://doi.org/10.1142/S0218271892000057} {\bibfield  {journal} {\bibinfo
  {journal} {Int. J. Mod. Phys. D}\ }\textbf {\bibinfo {volume} {1}},\ \bibinfo
  {pages} {101} (\bibinfo {year} {1992})}\BibitemShut {NoStop}%
\bibitem [{\citenamefont {Jetzer}(1992)}]{Jetzer:1991jr}%
  \BibitemOpen
  \bibfield  {author} {\bibinfo {author} {\bibfnamefont {P.}~\bibnamefont
  {Jetzer}},\ }\bibfield  {title} {\bibinfo {title} {{Boson stars}},\ }\href
  {https://doi.org/10.1016/0370-1573(92)90123-H} {\bibfield  {journal}
  {\bibinfo  {journal} {Phys. Rept.}\ }\textbf {\bibinfo {volume} {220}},\
  \bibinfo {pages} {163} (\bibinfo {year} {1992})}\BibitemShut {NoStop}%
\bibitem [{\citenamefont {Mielke}\ and\ \citenamefont
  {Schunck}(1997)}]{Mielke:1997re}%
  \BibitemOpen
  \bibfield  {author} {\bibinfo {author} {\bibfnamefont {E.~W.}\ \bibnamefont
  {Mielke}}\ and\ \bibinfo {author} {\bibfnamefont {F.~E.}\ \bibnamefont
  {Schunck}},\ }\bibfield  {title} {\bibinfo {title} {{Boson stars: Early
  history and recent prospects}},\ }in\ \href@noop {} {\emph {\bibinfo
  {booktitle} {{8th Marcel Grossmann Meeting on Recent Developments in
  Theoretical and Experimental General Relativity, Gravitation and Relativistic
  Field Theories (MG 8)}}}}\ (\bibinfo {year} {1997})\ pp.\ \bibinfo {pages}
  {1607--1626},\ \Eprint {https://arxiv.org/abs/gr-qc/9801063}
  {arXiv:gr-qc/9801063} \BibitemShut {NoStop}%
\bibitem [{\citenamefont {Schunck}\ and\ \citenamefont
  {Mielke}(2003)}]{Schunck:2003kk}%
  \BibitemOpen
  \bibfield  {author} {\bibinfo {author} {\bibfnamefont {F.~E.}\ \bibnamefont
  {Schunck}}\ and\ \bibinfo {author} {\bibfnamefont {E.~W.}\ \bibnamefont
  {Mielke}},\ }\bibfield  {title} {\bibinfo {title} {{General relativistic
  boson stars}},\ }\href {https://doi.org/10.1088/0264-9381/20/20/201}
  {\bibfield  {journal} {\bibinfo  {journal} {Class. Quant. Grav.}\ }\textbf
  {\bibinfo {volume} {20}},\ \bibinfo {pages} {R301} (\bibinfo {year}
  {2003})},\ \Eprint {https://arxiv.org/abs/0801.0307} {arXiv:0801.0307
  [astro-ph]} \BibitemShut {NoStop}%
\bibitem [{\citenamefont {Mielke}(2016)}]{Mielke:2016war}%
  \BibitemOpen
  \bibfield  {author} {\bibinfo {author} {\bibfnamefont {E.~W.}\ \bibnamefont
  {Mielke}},\ }\bibfield  {title} {\bibinfo {title} {{Rotating Boson Stars}},\
  }\href {https://doi.org/10.1007/978-3-319-31299-6_6} {\bibfield  {journal}
  {\bibinfo  {journal} {Fundam. Theor. Phys.}\ }\textbf {\bibinfo {volume}
  {183}},\ \bibinfo {pages} {115} (\bibinfo {year} {2016})}\BibitemShut
  {NoStop}%
\bibitem [{\citenamefont {Visinelli}(2021)}]{Visinelli:2021uve}%
  \BibitemOpen
  \bibfield  {author} {\bibinfo {author} {\bibfnamefont {L.}~\bibnamefont
  {Visinelli}},\ }\bibfield  {title} {\bibinfo {title} {{Boson stars and
  oscillatons: A review}},\ }\href {https://doi.org/10.1142/S0218271821300068}
  {\bibfield  {journal} {\bibinfo  {journal} {Int. J. Mod. Phys. D}\ }\textbf
  {\bibinfo {volume} {30}},\ \bibinfo {pages} {2130006} (\bibinfo {year}
  {2021})},\ \Eprint {https://arxiv.org/abs/2109.05481} {arXiv:2109.05481
  [gr-qc]} \BibitemShut {NoStop}%
\bibitem [{\citenamefont {Shnir}(2023)}]{Shnir:2022lba}%
  \BibitemOpen
  \bibfield  {author} {\bibinfo {author} {\bibfnamefont {Y.}~\bibnamefont
  {Shnir}},\ }\bibfield  {title} {\bibinfo {title} {{Boson Stars}},\ }\href
  {https://doi.org/10.1007/978-3-031-31520-6_10} {\bibfield  {journal}
  {\bibinfo  {journal} {Lect. Notes Phys.}\ }\textbf {\bibinfo {volume}
  {1017}},\ \bibinfo {pages} {347} (\bibinfo {year} {2023})},\ \Eprint
  {https://arxiv.org/abs/2204.06374} {arXiv:2204.06374 [gr-qc]} \BibitemShut
  {NoStop}%
\bibitem [{\citenamefont {Liebling}\ and\ \citenamefont
  {Palenzuela}(2023)}]{Liebling:2012fv}%
  \BibitemOpen
  \bibfield  {author} {\bibinfo {author} {\bibfnamefont {S.~L.}\ \bibnamefont
  {Liebling}}\ and\ \bibinfo {author} {\bibfnamefont {C.}~\bibnamefont
  {Palenzuela}},\ }\bibfield  {title} {\bibinfo {title} {{Dynamical boson
  stars}},\ }\href {https://doi.org/10.1007/s41114-023-00043-4} {\bibfield
  {journal} {\bibinfo  {journal} {Living Rev. Rel.}\ }\textbf {\bibinfo
  {volume} {26}},\ \bibinfo {pages} {1} (\bibinfo {year} {2023})},\ \Eprint
  {https://arxiv.org/abs/1202.5809} {arXiv:1202.5809 [gr-qc]} \BibitemShut
  {NoStop}%
\bibitem [{\citenamefont {Sin}(1994)}]{Sin:1992bg}%
  \BibitemOpen
  \bibfield  {author} {\bibinfo {author} {\bibfnamefont {S.-J.}\ \bibnamefont
  {Sin}},\ }\bibfield  {title} {\bibinfo {title} {{Late time cosmological phase
  transition and galactic halo as Bose liquid}},\ }\href
  {https://doi.org/10.1103/PhysRevD.50.3650} {\bibfield  {journal} {\bibinfo
  {journal} {Phys. Rev. D}\ }\textbf {\bibinfo {volume} {50}},\ \bibinfo
  {pages} {3650} (\bibinfo {year} {1994})},\ \Eprint
  {https://arxiv.org/abs/hep-ph/9205208} {arXiv:hep-ph/9205208} \BibitemShut
  {NoStop}%
\bibitem [{\citenamefont {Schive}\ \emph {et~al.}(2014)\citenamefont {Schive},
  \citenamefont {Chiueh},\ and\ \citenamefont {Broadhurst}}]{Schive:2014dra}%
  \BibitemOpen
  \bibfield  {author} {\bibinfo {author} {\bibfnamefont {H.-Y.}\ \bibnamefont
  {Schive}}, \bibinfo {author} {\bibfnamefont {T.}~\bibnamefont {Chiueh}},\
  and\ \bibinfo {author} {\bibfnamefont {T.}~\bibnamefont {Broadhurst}},\
  }\bibfield  {title} {\bibinfo {title} {{Cosmic Structure as the Quantum
  Interference of a Coherent Dark Wave}},\ }\href
  {https://doi.org/10.1038/nphys2996} {\bibfield  {journal} {\bibinfo
  {journal} {Nature Phys.}\ }\textbf {\bibinfo {volume} {10}},\ \bibinfo
  {pages} {496} (\bibinfo {year} {2014})},\ \Eprint
  {https://arxiv.org/abs/1406.6586} {arXiv:1406.6586 [astro-ph.GA]}
  \BibitemShut {NoStop}%
\bibitem [{\citenamefont {Alcubierre}\ \emph {et~al.}(2025)\citenamefont
  {Alcubierre}, \citenamefont {Barranco}, \citenamefont {Bernal}, \citenamefont
  {Degollado}, \citenamefont {Diez-Tejedor}, \citenamefont {Megevand},
  \citenamefont {N{\'u}{\~n}ez},\ and\ \citenamefont
  {Sarbach}}]{Alcubierre:2024mtq}%
  \BibitemOpen
  \bibfield  {author} {\bibinfo {author} {\bibfnamefont {M.}~\bibnamefont
  {Alcubierre}}, \bibinfo {author} {\bibfnamefont {J.}~\bibnamefont
  {Barranco}}, \bibinfo {author} {\bibfnamefont {A.}~\bibnamefont {Bernal}},
  \bibinfo {author} {\bibfnamefont {J.~C.}\ \bibnamefont {Degollado}}, \bibinfo
  {author} {\bibfnamefont {A.}~\bibnamefont {Diez-Tejedor}}, \bibinfo {author}
  {\bibfnamefont {M.}~\bibnamefont {Megevand}}, \bibinfo {author}
  {\bibfnamefont {D.}~\bibnamefont {N{\'u}{\~n}ez}},\ and\ \bibinfo {author}
  {\bibfnamefont {O.}~\bibnamefont {Sarbach}},\ }\bibfield  {title} {\bibinfo
  {title} {{Gravitational atoms beyond the test field limit: the case of Sgr A*
  and ultralight dark matter}},\ }\href
  {https://doi.org/10.1088/1361-6382/ae110a} {\bibfield  {journal} {\bibinfo
  {journal} {Class. Quant. Grav.}\ }\textbf {\bibinfo {volume} {42}},\ \bibinfo
  {pages} {21LT01} (\bibinfo {year} {2025})},\ \Eprint
  {https://arxiv.org/abs/2411.18601} {arXiv:2411.18601 [gr-qc]} \BibitemShut
  {NoStop}%
\bibitem [{\citenamefont {Alcubierre}\ \emph {et~al.}(2026)\citenamefont
  {Alcubierre}, \citenamefont {Barranco}, \citenamefont {Bernal}, \citenamefont
  {Degollado}, \citenamefont {Diez-Tejedor}, \citenamefont {Megevand},
  \citenamefont {Nunez},\ and\ \citenamefont {Sarbach}}]{Alcubierre:2025zus}%
  \BibitemOpen
  \bibfield  {author} {\bibinfo {author} {\bibfnamefont {M.}~\bibnamefont
  {Alcubierre}}, \bibinfo {author} {\bibfnamefont {J.}~\bibnamefont
  {Barranco}}, \bibinfo {author} {\bibfnamefont {A.}~\bibnamefont {Bernal}},
  \bibinfo {author} {\bibfnamefont {J.~C.}\ \bibnamefont {Degollado}}, \bibinfo
  {author} {\bibfnamefont {A.}~\bibnamefont {Diez-Tejedor}}, \bibinfo {author}
  {\bibfnamefont {M.}~\bibnamefont {Megevand}}, \bibinfo {author}
  {\bibfnamefont {D.}~\bibnamefont {Nunez}},\ and\ \bibinfo {author}
  {\bibfnamefont {O.}~\bibnamefont {Sarbach}},\ }\bibfield  {title} {\bibinfo
  {title} {{Noble gravitational atoms: self-gravitating black hole scalar wigs
  with angular momentum number}},\ }\href
  {https://doi.org/10.1088/1361-6382/ae4160} {\bibfield  {journal} {\bibinfo
  {journal} {Class. Quant. Grav.}\ }\textbf {\bibinfo {volume} {43}},\ \bibinfo
  {pages} {045010} (\bibinfo {year} {2026})},\ \Eprint
  {https://arxiv.org/abs/2512.08095} {arXiv:2512.08095 [gr-qc]} \BibitemShut
  {NoStop}%
\bibitem [{\citenamefont {Visinelli}\ \emph {et~al.}(2018)\citenamefont
  {Visinelli}, \citenamefont {Baum}, \citenamefont {Redondo}, \citenamefont
  {Freese},\ and\ \citenamefont {Wilczek}}]{Visinelli:2017ooc}%
  \BibitemOpen
  \bibfield  {author} {\bibinfo {author} {\bibfnamefont {L.}~\bibnamefont
  {Visinelli}}, \bibinfo {author} {\bibfnamefont {S.}~\bibnamefont {Baum}},
  \bibinfo {author} {\bibfnamefont {J.}~\bibnamefont {Redondo}}, \bibinfo
  {author} {\bibfnamefont {K.}~\bibnamefont {Freese}},\ and\ \bibinfo {author}
  {\bibfnamefont {F.}~\bibnamefont {Wilczek}},\ }\bibfield  {title} {\bibinfo
  {title} {{Dilute and dense axion stars}},\ }\href
  {https://doi.org/10.1016/j.physletb.2017.12.010} {\bibfield  {journal}
  {\bibinfo  {journal} {Phys. Lett. B}\ }\textbf {\bibinfo {volume} {777}},\
  \bibinfo {pages} {64} (\bibinfo {year} {2018})},\ \Eprint
  {https://arxiv.org/abs/1710.08910} {arXiv:1710.08910 [astro-ph.CO]}
  \BibitemShut {NoStop}%
\bibitem [{\citenamefont {Braaten}\ \emph {et~al.}(2016)\citenamefont
  {Braaten}, \citenamefont {Mohapatra},\ and\ \citenamefont
  {Zhang}}]{Braaten:2015eeu}%
  \BibitemOpen
  \bibfield  {author} {\bibinfo {author} {\bibfnamefont {E.}~\bibnamefont
  {Braaten}}, \bibinfo {author} {\bibfnamefont {A.}~\bibnamefont {Mohapatra}},\
  and\ \bibinfo {author} {\bibfnamefont {H.}~\bibnamefont {Zhang}},\ }\bibfield
   {title} {\bibinfo {title} {{Dense Axion Stars}},\ }\href
  {https://doi.org/10.1103/PhysRevLett.117.121801} {\bibfield  {journal}
  {\bibinfo  {journal} {Phys. Rev. Lett.}\ }\textbf {\bibinfo {volume} {117}},\
  \bibinfo {pages} {121801} (\bibinfo {year} {2016})},\ \Eprint
  {https://arxiv.org/abs/1512.00108} {arXiv:1512.00108 [hep-ph]} \BibitemShut
  {NoStop}%
\bibitem [{\citenamefont {Guerra}\ \emph {et~al.}(2019)\citenamefont {Guerra},
  \citenamefont {Macedo},\ and\ \citenamefont {Pani}}]{Guerra:2019srj}%
  \BibitemOpen
  \bibfield  {author} {\bibinfo {author} {\bibfnamefont {D.}~\bibnamefont
  {Guerra}}, \bibinfo {author} {\bibfnamefont {C.~F.~B.}\ \bibnamefont
  {Macedo}},\ and\ \bibinfo {author} {\bibfnamefont {P.}~\bibnamefont {Pani}},\
  }\bibfield  {title} {\bibinfo {title} {{Axion boson stars}},\ }\href
  {https://doi.org/10.1088/1475-7516/2019/09/061} {\bibfield  {journal}
  {\bibinfo  {journal} {JCAP}\ }\textbf {\bibinfo {volume} {09}}\bibfield
  {number} {\bibinfo  {number} { (09)},\ \bibinfo {pages} {061}},\ }\bibinfo
  {note} {[Erratum: JCAP 06, E01 (2020)]},\ \Eprint
  {https://arxiv.org/abs/1909.05515} {arXiv:1909.05515 [gr-qc]} \BibitemShut
  {NoStop}%
\bibitem [{\citenamefont {Peccei}\ and\ \citenamefont
  {Quinn}(1977)}]{Peccei:1977hh}%
  \BibitemOpen
  \bibfield  {author} {\bibinfo {author} {\bibfnamefont {R.~D.}\ \bibnamefont
  {Peccei}}\ and\ \bibinfo {author} {\bibfnamefont {H.~R.}\ \bibnamefont
  {Quinn}},\ }\bibfield  {title} {\bibinfo {title} {{CP Conservation in the
  Presence of Instantons}},\ }\href
  {https://doi.org/10.1103/PhysRevLett.38.1440} {\bibfield  {journal} {\bibinfo
   {journal} {Phys. Rev. Lett.}\ }\textbf {\bibinfo {volume} {38}},\ \bibinfo
  {pages} {1440} (\bibinfo {year} {1977})}\BibitemShut {NoStop}%
\bibitem [{\citenamefont {Wilczek}(1978)}]{Wilczek:1977pj}%
  \BibitemOpen
  \bibfield  {author} {\bibinfo {author} {\bibfnamefont {F.}~\bibnamefont
  {Wilczek}},\ }\bibfield  {title} {\bibinfo {title} {{Problem of Strong $P$
  and $T$ Invariance in the Presence of Instantons}},\ }\href
  {https://doi.org/10.1103/PhysRevLett.40.279} {\bibfield  {journal} {\bibinfo
  {journal} {Phys. Rev. Lett.}\ }\textbf {\bibinfo {volume} {40}},\ \bibinfo
  {pages} {279} (\bibinfo {year} {1978})}\BibitemShut {NoStop}%
\bibitem [{\citenamefont {Weinberg}(1978)}]{Weinberg:1977ma}%
  \BibitemOpen
  \bibfield  {author} {\bibinfo {author} {\bibfnamefont {S.}~\bibnamefont
  {Weinberg}},\ }\bibfield  {title} {\bibinfo {title} {{A New Light Boson?}},\
  }\href {https://doi.org/10.1103/PhysRevLett.40.223} {\bibfield  {journal}
  {\bibinfo  {journal} {Phys. Rev. Lett.}\ }\textbf {\bibinfo {volume} {40}},\
  \bibinfo {pages} {223} (\bibinfo {year} {1978})}\BibitemShut {NoStop}%
\bibitem [{\citenamefont {Sikivie}(2024)}]{Sikivie:2024isv}%
  \BibitemOpen
  \bibfield  {author} {\bibinfo {author} {\bibfnamefont {P.}~\bibnamefont
  {Sikivie}},\ }\bibfield  {title} {\bibinfo {title} {{Axion dark matter}},\
  }\href {https://doi.org/10.1016/j.nuclphysb.2024.116500} {\bibfield
  {journal} {\bibinfo  {journal} {Nucl. Phys. B}\ }\textbf {\bibinfo {volume}
  {1003}},\ \bibinfo {pages} {116500} (\bibinfo {year} {2024})}\BibitemShut
  {NoStop}%
\bibitem [{\citenamefont {Torres}\ \emph {et~al.}(2000)\citenamefont {Torres},
  \citenamefont {Capozziello},\ and\ \citenamefont {Lambiase}}]{Torres:2000dw}%
  \BibitemOpen
  \bibfield  {author} {\bibinfo {author} {\bibfnamefont {D.~F.}\ \bibnamefont
  {Torres}}, \bibinfo {author} {\bibfnamefont {S.}~\bibnamefont
  {Capozziello}},\ and\ \bibinfo {author} {\bibfnamefont {G.}~\bibnamefont
  {Lambiase}},\ }\bibfield  {title} {\bibinfo {title} {{A Supermassive scalar
  star at the galactic center?}},\ }\href
  {https://doi.org/10.1103/PhysRevD.62.104012} {\bibfield  {journal} {\bibinfo
  {journal} {Phys. Rev. D}\ }\textbf {\bibinfo {volume} {62}},\ \bibinfo
  {pages} {104012} (\bibinfo {year} {2000})},\ \Eprint
  {https://arxiv.org/abs/astro-ph/0004064} {arXiv:astro-ph/0004064}
  \BibitemShut {NoStop}%
\bibitem [{\citenamefont {Guzman}(2006)}]{Guzman:2005bs}%
  \BibitemOpen
  \bibfield  {author} {\bibinfo {author} {\bibfnamefont {F.~S.}\ \bibnamefont
  {Guzman}},\ }\bibfield  {title} {\bibinfo {title} {{Accretion disc onto boson
  stars: A Way to supplant black holes candidates}},\ }\href
  {https://doi.org/10.1103/PhysRevD.73.021501} {\bibfield  {journal} {\bibinfo
  {journal} {Phys. Rev. D}\ }\textbf {\bibinfo {volume} {73}},\ \bibinfo
  {pages} {021501} (\bibinfo {year} {2006})},\ \Eprint
  {https://arxiv.org/abs/gr-qc/0512081} {arXiv:gr-qc/0512081} \BibitemShut
  {NoStop}%
\bibitem [{\citenamefont {Amaro-Seoane}\ \emph {et~al.}(2010)\citenamefont
  {Amaro-Seoane}, \citenamefont {Barranco}, \citenamefont {Bernal},\ and\
  \citenamefont {Rezzolla}}]{Amaro-Seoane:2010pks}%
  \BibitemOpen
  \bibfield  {author} {\bibinfo {author} {\bibfnamefont {P.}~\bibnamefont
  {Amaro-Seoane}}, \bibinfo {author} {\bibfnamefont {J.}~\bibnamefont
  {Barranco}}, \bibinfo {author} {\bibfnamefont {A.}~\bibnamefont {Bernal}},\
  and\ \bibinfo {author} {\bibfnamefont {L.}~\bibnamefont {Rezzolla}},\
  }\bibfield  {title} {\bibinfo {title} {{Constraining scalar fields with
  stellar kinematics and collisional dark matter}},\ }\href
  {https://doi.org/10.1088/1475-7516/2010/11/002} {\bibfield  {journal}
  {\bibinfo  {journal} {JCAP}\ }\textbf {\bibinfo {volume} {11}},\ \bibinfo
  {pages} {002}},\ \Eprint {https://arxiv.org/abs/1009.0019} {arXiv:1009.0019
  [astro-ph.CO]} \BibitemShut {NoStop}%
\bibitem [{\citenamefont {Cardoso}\ and\ \citenamefont
  {Pani}(2019)}]{Cardoso:2019rvt}%
  \BibitemOpen
  \bibfield  {author} {\bibinfo {author} {\bibfnamefont {V.}~\bibnamefont
  {Cardoso}}\ and\ \bibinfo {author} {\bibfnamefont {P.}~\bibnamefont {Pani}},\
  }\bibfield  {title} {\bibinfo {title} {{Testing the nature of dark compact
  objects: a status report}},\ }\href
  {https://doi.org/10.1007/s41114-019-0020-4} {\bibfield  {journal} {\bibinfo
  {journal} {Living Rev. Rel.}\ }\textbf {\bibinfo {volume} {22}},\ \bibinfo
  {pages} {4} (\bibinfo {year} {2019})},\ \Eprint
  {https://arxiv.org/abs/1904.05363} {arXiv:1904.05363 [gr-qc]} \BibitemShut
  {NoStop}%
\bibitem [{\citenamefont {Marks}\ \emph {et~al.}(2025)\citenamefont {Marks},
  \citenamefont {Staelens}, \citenamefont {Evstafyeva},\ and\ \citenamefont
  {Sperhake}}]{Marks:2025jpt}%
  \BibitemOpen
  \bibfield  {author} {\bibinfo {author} {\bibfnamefont {G.~A.}\ \bibnamefont
  {Marks}}, \bibinfo {author} {\bibfnamefont {S.~J.}\ \bibnamefont {Staelens}},
  \bibinfo {author} {\bibfnamefont {T.}~\bibnamefont {Evstafyeva}},\ and\
  \bibinfo {author} {\bibfnamefont {U.}~\bibnamefont {Sperhake}},\ }\bibfield
  {title} {\bibinfo {title} {{Long-Term Stable Nonlinear Evolutions of
  Ultracompact Black-Hole Mimickers}},\ }\href
  {https://doi.org/10.1103/lk48-7r2f} {\bibfield  {journal} {\bibinfo
  {journal} {Phys. Rev. Lett.}\ }\textbf {\bibinfo {volume} {135}},\ \bibinfo
  {pages} {131402} (\bibinfo {year} {2025})},\ \Eprint
  {https://arxiv.org/abs/2504.17775} {arXiv:2504.17775 [gr-qc]} \BibitemShut
  {NoStop}%
\bibitem [{\citenamefont {Astefanesei}\ and\ \citenamefont
  {Radu}(2003)}]{Astefanesei:2003qy}%
  \BibitemOpen
  \bibfield  {author} {\bibinfo {author} {\bibfnamefont {D.}~\bibnamefont
  {Astefanesei}}\ and\ \bibinfo {author} {\bibfnamefont {E.}~\bibnamefont
  {Radu}},\ }\bibfield  {title} {\bibinfo {title} {{Boson stars with negative
  cosmological constant}},\ }\href
  {https://doi.org/10.1016/S0550-3213(03)00482-6} {\bibfield  {journal}
  {\bibinfo  {journal} {Nucl. Phys. B}\ }\textbf {\bibinfo {volume} {665}},\
  \bibinfo {pages} {594} (\bibinfo {year} {2003})},\ \Eprint
  {https://arxiv.org/abs/gr-qc/0309131} {arXiv:gr-qc/0309131} \BibitemShut
  {NoStop}%
\bibitem [{\citenamefont {Radu}\ and\ \citenamefont
  {Subagyo}(2012)}]{Radu:2012yx}%
  \BibitemOpen
  \bibfield  {author} {\bibinfo {author} {\bibfnamefont {E.}~\bibnamefont
  {Radu}}\ and\ \bibinfo {author} {\bibfnamefont {B.}~\bibnamefont {Subagyo}},\
  }\bibfield  {title} {\bibinfo {title} {{Spinning scalar solitons in anti-de
  Sitter spacetime}},\ }\href {https://doi.org/10.1016/j.physletb.2012.09.050}
  {\bibfield  {journal} {\bibinfo  {journal} {Phys. Lett. B}\ }\textbf
  {\bibinfo {volume} {717}},\ \bibinfo {pages} {450} (\bibinfo {year}
  {2012})},\ \Eprint {https://arxiv.org/abs/1207.3715} {arXiv:1207.3715
  [gr-qc]} \BibitemShut {NoStop}%
\bibitem [{\citenamefont {Buchel}\ \emph {et~al.}(2013)\citenamefont {Buchel},
  \citenamefont {Liebling},\ and\ \citenamefont {Lehner}}]{Buchel:2013uba}%
  \BibitemOpen
  \bibfield  {author} {\bibinfo {author} {\bibfnamefont {A.}~\bibnamefont
  {Buchel}}, \bibinfo {author} {\bibfnamefont {S.~L.}\ \bibnamefont
  {Liebling}},\ and\ \bibinfo {author} {\bibfnamefont {L.}~\bibnamefont
  {Lehner}},\ }\bibfield  {title} {\bibinfo {title} {{Boson stars in AdS
  spacetime}},\ }\href {https://doi.org/10.1103/PhysRevD.87.123006} {\bibfield
  {journal} {\bibinfo  {journal} {Phys. Rev. D}\ }\textbf {\bibinfo {volume}
  {87}},\ \bibinfo {pages} {123006} (\bibinfo {year} {2013})},\ \Eprint
  {https://arxiv.org/abs/1304.4166} {arXiv:1304.4166 [gr-qc]} \BibitemShut
  {NoStop}%
\bibitem [{\citenamefont {Maliborski}\ and\ \citenamefont
  {Rostworowski}(2013)}]{Maliborski:2013ula}%
  \BibitemOpen
  \bibfield  {author} {\bibinfo {author} {\bibfnamefont {M.}~\bibnamefont
  {Maliborski}}\ and\ \bibinfo {author} {\bibfnamefont {A.}~\bibnamefont
  {Rostworowski}},\ }\bibfield  {title} {\bibinfo {title} {{A comment on
  ''Boson stars in AdS''}},\ }\href@noop {} {\bibfield  {journal} {\bibinfo
  {journal} {arXive}\ } (\bibinfo {year} {2013})},\ \Eprint
  {https://arxiv.org/abs/1307.2875} {arXiv:1307.2875 [gr-qc]} \BibitemShut
  {NoStop}%
\bibitem [{\citenamefont {Brihaye}\ \emph {et~al.}(2015)\citenamefont
  {Brihaye}, \citenamefont {Hartmann},\ and\ \citenamefont
  {Riedel}}]{Brihaye:2014bqa}%
  \BibitemOpen
  \bibfield  {author} {\bibinfo {author} {\bibfnamefont {Y.}~\bibnamefont
  {Brihaye}}, \bibinfo {author} {\bibfnamefont {B.}~\bibnamefont {Hartmann}},\
  and\ \bibinfo {author} {\bibfnamefont {J.}~\bibnamefont {Riedel}},\
  }\bibfield  {title} {\bibinfo {title} {{Self-interacting boson stars with a
  single Killing vector field in anti{\textendash}de Sitter space-time}},\
  }\href {https://doi.org/10.1103/PhysRevD.92.044049} {\bibfield  {journal}
  {\bibinfo  {journal} {Phys. Rev. D}\ }\textbf {\bibinfo {volume} {92}},\
  \bibinfo {pages} {044049} (\bibinfo {year} {2015})},\ \Eprint
  {https://arxiv.org/abs/1404.1874} {arXiv:1404.1874 [gr-qc]} \BibitemShut
  {NoStop}%
\bibitem [{\citenamefont {Fodor}\ \emph {et~al.}(2015)\citenamefont {Fodor},
  \citenamefont {Forg{\'a}cs},\ and\ \citenamefont
  {Grandcl{\'e}ment}}]{Fodor:2015eia}%
  \BibitemOpen
  \bibfield  {author} {\bibinfo {author} {\bibfnamefont {G.}~\bibnamefont
  {Fodor}}, \bibinfo {author} {\bibfnamefont {P.}~\bibnamefont {Forg{\'a}cs}},\
  and\ \bibinfo {author} {\bibfnamefont {P.}~\bibnamefont {Grandcl{\'e}ment}},\
  }\bibfield  {title} {\bibinfo {title} {{Self-gravitating scalar breathers
  with negative cosmological constant}},\ }\href
  {https://doi.org/10.1103/PhysRevD.92.025036} {\bibfield  {journal} {\bibinfo
  {journal} {Phys. Rev. D}\ }\textbf {\bibinfo {volume} {92}},\ \bibinfo
  {pages} {025036} (\bibinfo {year} {2015})},\ \Eprint
  {https://arxiv.org/abs/1503.07746} {arXiv:1503.07746 [gr-qc]} \BibitemShut
  {NoStop}%
\bibitem [{\citenamefont {Chru{\'s}ciel}\ \emph {et~al.}(2018)\citenamefont
  {Chru{\'s}ciel}, \citenamefont {Delay}, \citenamefont {Klinger},
  \citenamefont {Kriegl}, \citenamefont {Michor},\ and\ \citenamefont
  {Rainer}}]{Chrusciel:2017uor}%
  \BibitemOpen
  \bibfield  {author} {\bibinfo {author} {\bibfnamefont {P.~T.}\ \bibnamefont
  {Chru{\'s}ciel}}, \bibinfo {author} {\bibfnamefont {E.}~\bibnamefont
  {Delay}}, \bibinfo {author} {\bibfnamefont {P.}~\bibnamefont {Klinger}},
  \bibinfo {author} {\bibfnamefont {A.}~\bibnamefont {Kriegl}}, \bibinfo
  {author} {\bibfnamefont {P.~W.}\ \bibnamefont {Michor}},\ and\ \bibinfo
  {author} {\bibfnamefont {A.}~\bibnamefont {Rainer}},\ }\bibfield  {title}
  {\bibinfo {title} {{Non-singular space-times with a negative cosmological
  constant: V. Boson stars}},\ }\href
  {https://doi.org/10.1007/s11005-018-1062-3} {\bibfield  {journal} {\bibinfo
  {journal} {Lett. Math. Phys.}\ }\textbf {\bibinfo {volume} {108}},\ \bibinfo
  {pages} {2009} (\bibinfo {year} {2018})},\ \Eprint
  {https://arxiv.org/abs/1708.02878} {arXiv:1708.02878 [gr-qc]} \BibitemShut
  {NoStop}%
\bibitem [{\citenamefont {Liu}\ \emph {et~al.}(2020)\citenamefont {Liu},
  \citenamefont {Lu},\ and\ \citenamefont {Pang}}]{Liu:2020uaz}%
  \BibitemOpen
  \bibfield  {author} {\bibinfo {author} {\bibfnamefont {H.-S.}\ \bibnamefont
  {Liu}}, \bibinfo {author} {\bibfnamefont {H.}~\bibnamefont {Lu}},\ and\
  \bibinfo {author} {\bibfnamefont {Y.}~\bibnamefont {Pang}},\ }\bibfield
  {title} {\bibinfo {title} {{Revisiting the AdS Boson Stars: the Mass-Charge
  Relations}},\ }\href {https://doi.org/10.1103/PhysRevD.102.126008} {\bibfield
   {journal} {\bibinfo  {journal} {Phys. Rev. D}\ }\textbf {\bibinfo {volume}
  {102}},\ \bibinfo {pages} {126008} (\bibinfo {year} {2020})},\ \Eprint
  {https://arxiv.org/abs/2007.15017} {arXiv:2007.15017 [hep-th]} \BibitemShut
  {NoStop}%
\bibitem [{\citenamefont {Guo}\ \emph {et~al.}(2021)\citenamefont {Guo},
  \citenamefont {Liu}, \citenamefont {L{\"u}},\ and\ \citenamefont
  {Pang}}]{Guo:2020bqz}%
  \BibitemOpen
  \bibfield  {author} {\bibinfo {author} {\bibfnamefont {S.-F.}\ \bibnamefont
  {Guo}}, \bibinfo {author} {\bibfnamefont {H.-S.}\ \bibnamefont {Liu}},
  \bibinfo {author} {\bibfnamefont {H.}~\bibnamefont {L{\"u}}},\ and\ \bibinfo
  {author} {\bibfnamefont {Y.}~\bibnamefont {Pang}},\ }\bibfield  {title}
  {\bibinfo {title} {{Large-charge limit of AdS boson stars with mixed boundary
  conditions}},\ }\href {https://doi.org/10.1007/JHEP04(2021)220} {\bibfield
  {journal} {\bibinfo  {journal} {JHEP}\ }\textbf {\bibinfo {volume} {04}},\
  \bibinfo {pages} {220}},\ \Eprint {https://arxiv.org/abs/2101.00017}
  {arXiv:2101.00017 [hep-th]} \BibitemShut {NoStop}%
\bibitem [{\citenamefont {Herdeiro}\ \emph {et~al.}(2024)\citenamefont
  {Herdeiro}, \citenamefont {Huang}, \citenamefont {Kunz},\ and\ \citenamefont
  {Radu}}]{Herdeiro:2024myz}%
  \BibitemOpen
  \bibfield  {author} {\bibinfo {author} {\bibfnamefont {C.}~\bibnamefont
  {Herdeiro}}, \bibinfo {author} {\bibfnamefont {H.}~\bibnamefont {Huang}},
  \bibinfo {author} {\bibfnamefont {J.}~\bibnamefont {Kunz}},\ and\ \bibinfo
  {author} {\bibfnamefont {E.}~\bibnamefont {Radu}},\ }\bibfield  {title}
  {\bibinfo {title} {{Einstein-(complex)-Maxwell static boson stars in AdS}},\
  }\href {https://doi.org/10.1016/j.physletb.2024.138939} {\bibfield  {journal}
  {\bibinfo  {journal} {Phys. Lett. B}\ }\textbf {\bibinfo {volume} {856}},\
  \bibinfo {pages} {138939} (\bibinfo {year} {2024})},\ \Eprint
  {https://arxiv.org/abs/2405.10671} {arXiv:2405.10671 [gr-qc]} \BibitemShut
  {NoStop}%
\bibitem [{\citenamefont {Liu}\ \emph {et~al.}(2025)\citenamefont {Liu},
  \citenamefont {Wang},\ and\ \citenamefont {Zhao}}]{Liu:2025fcs}%
  \BibitemOpen
  \bibfield  {author} {\bibinfo {author} {\bibfnamefont {S.-C.}\ \bibnamefont
  {Liu}}, \bibinfo {author} {\bibfnamefont {Y.-Q.}\ \bibnamefont {Wang}},\ and\
  \bibinfo {author} {\bibfnamefont {Z.-H.}\ \bibnamefont {Zhao}},\ }\bibfield
  {title} {\bibinfo {title} {{Frozen solitonic Hayward-boson stars in Anti-de
  Sitter Spacetime}},\ }\href@noop {} {\  (\bibinfo {year} {2025})},\ \Eprint
  {https://arxiv.org/abs/2512.10197} {arXiv:2512.10197 [gr-qc]} \BibitemShut
  {NoStop}%
\bibitem [{\citenamefont {Zhao}\ \emph {et~al.}(2025)\citenamefont {Zhao},
  \citenamefont {Gu}, \citenamefont {Liu}, \citenamefont {Liu},\ and\
  \citenamefont {Wang}}]{Zhao:2025yhy}%
  \BibitemOpen
  \bibfield  {author} {\bibinfo {author} {\bibfnamefont {Z.-H.}\ \bibnamefont
  {Zhao}}, \bibinfo {author} {\bibfnamefont {Y.-N.}\ \bibnamefont {Gu}},
  \bibinfo {author} {\bibfnamefont {S.-C.}\ \bibnamefont {Liu}}, \bibinfo
  {author} {\bibfnamefont {Z.-Q.}\ \bibnamefont {Liu}},\ and\ \bibinfo {author}
  {\bibfnamefont {Y.-Q.}\ \bibnamefont {Wang}},\ }\bibfield  {title} {\bibinfo
  {title} {{Light Rings, Accretion Disks and Shadows of Hayward Boson Stars in
  asymptotically AdS Spacetime}},\ }\href@noop {} {\  (\bibinfo {year}
  {2025})},\ \Eprint {https://arxiv.org/abs/2507.09563} {arXiv:2507.09563
  [gr-qc]} \BibitemShut {NoStop}%
\bibitem [{\citenamefont {Hawking}\ and\ \citenamefont
  {Ellis}(1973)}]{hawking1975large}%
  \BibitemOpen
  \bibfield  {author} {\bibinfo {author} {\bibfnamefont {S.~W.}\ \bibnamefont
  {Hawking}}\ and\ \bibinfo {author} {\bibfnamefont {G.~F.~R.}\ \bibnamefont
  {Ellis}},\ }\href@noop {} {\emph {\bibinfo {title} {The Large Scale Structure
  of Space-Time}}},\ Cambridge Monographs on Mathematical Physics\ (\bibinfo
  {publisher} {Cambridge University Press},\ \bibinfo {year}
  {1973})\BibitemShut {NoStop}%
\bibitem [{\citenamefont {Bosch}\ \emph {et~al.}(2016)\citenamefont {Bosch},
  \citenamefont {Green},\ and\ \citenamefont {Lehner}}]{Bosch:2016vcp}%
  \BibitemOpen
  \bibfield  {author} {\bibinfo {author} {\bibfnamefont {P.}~\bibnamefont
  {Bosch}}, \bibinfo {author} {\bibfnamefont {S.~R.}\ \bibnamefont {Green}},\
  and\ \bibinfo {author} {\bibfnamefont {L.}~\bibnamefont {Lehner}},\
  }\bibfield  {title} {\bibinfo {title} {{Nonlinear Evolution and Final Fate of
  Charged Anti{\textendash}de Sitter Black Hole Superradiant Instability}},\
  }\href {https://doi.org/10.1103/PhysRevLett.116.141102} {\bibfield  {journal}
  {\bibinfo  {journal} {Phys. Rev. Lett.}\ }\textbf {\bibinfo {volume} {116}},\
  \bibinfo {pages} {141102} (\bibinfo {year} {2016})},\ \Eprint
  {https://arxiv.org/abs/1601.01384} {arXiv:1601.01384 [gr-qc]} \BibitemShut
  {NoStop}%
\bibitem [{\citenamefont {Basu}\ \emph {et~al.}(2016)\citenamefont {Basu},
  \citenamefont {Krishnan},\ and\ \citenamefont
  {Bala~Subramanian}}]{Basu:2016srp}%
  \BibitemOpen
  \bibfield  {author} {\bibinfo {author} {\bibfnamefont {P.}~\bibnamefont
  {Basu}}, \bibinfo {author} {\bibfnamefont {C.}~\bibnamefont {Krishnan}},\
  and\ \bibinfo {author} {\bibfnamefont {P.~N.}\ \bibnamefont
  {Bala~Subramanian}},\ }\bibfield  {title} {\bibinfo {title} {{Hairy Black
  Holes in a Box}},\ }\href {https://doi.org/10.1007/JHEP11(2016)041}
  {\bibfield  {journal} {\bibinfo  {journal} {JHEP}\ }\textbf {\bibinfo
  {volume} {11}},\ \bibinfo {pages} {041}},\ \Eprint
  {https://arxiv.org/abs/1609.01208} {arXiv:1609.01208 [hep-th]} \BibitemShut
  {NoStop}%
\bibitem [{\citenamefont {Peng}(2017)}]{Peng:2017squ}%
  \BibitemOpen
  \bibfield  {author} {\bibinfo {author} {\bibfnamefont {Y.}~\bibnamefont
  {Peng}},\ }\bibfield  {title} {\bibinfo {title} {{Studies of a general flat
  space/boson star transition model in a box through a language similar to
  holographic superconductors}},\ }\href
  {https://doi.org/10.1007/JHEP07(2017)042} {\bibfield  {journal} {\bibinfo
  {journal} {JHEP}\ }\textbf {\bibinfo {volume} {07}},\ \bibinfo {pages}
  {042}},\ \Eprint {https://arxiv.org/abs/1705.08694} {arXiv:1705.08694
  [hep-th]} \BibitemShut {NoStop}%
\bibitem [{\citenamefont {Peng}\ \emph {et~al.}(2018)\citenamefont {Peng},
  \citenamefont {Wang},\ and\ \citenamefont {Liu}}]{Peng:2017gss}%
  \BibitemOpen
  \bibfield  {author} {\bibinfo {author} {\bibfnamefont {Y.}~\bibnamefont
  {Peng}}, \bibinfo {author} {\bibfnamefont {B.}~\bibnamefont {Wang}},\ and\
  \bibinfo {author} {\bibfnamefont {Y.}~\bibnamefont {Liu}},\ }\bibfield
  {title} {\bibinfo {title} {{On the thermodynamics of the black hole and hairy
  black hole transitions in the asymptotically flat spacetime with a box}},\
  }\href {https://doi.org/10.1140/epjc/s10052-018-5652-0} {\bibfield  {journal}
  {\bibinfo  {journal} {Eur. Phys. J. C}\ }\textbf {\bibinfo {volume} {78}},\
  \bibinfo {pages} {176} (\bibinfo {year} {2018})},\ \Eprint
  {https://arxiv.org/abs/1708.01411} {arXiv:1708.01411 [hep-th]} \BibitemShut
  {NoStop}%
\bibitem [{\citenamefont {Ferreira}\ and\ \citenamefont
  {Herdeiro}(2018)}]{Ferreira:2017tnc}%
  \BibitemOpen
  \bibfield  {author} {\bibinfo {author} {\bibfnamefont {H.~R.~C.}\
  \bibnamefont {Ferreira}}\ and\ \bibinfo {author} {\bibfnamefont {C.~A.~R.}\
  \bibnamefont {Herdeiro}},\ }\bibfield  {title} {\bibinfo {title}
  {{Superradiant instabilities in the Kerr-mirror and Kerr-AdS black holes with
  Robin boundary conditions}},\ }\href
  {https://doi.org/10.1103/PhysRevD.97.084003} {\bibfield  {journal} {\bibinfo
  {journal} {Phys. Rev. D}\ }\textbf {\bibinfo {volume} {97}},\ \bibinfo
  {pages} {084003} (\bibinfo {year} {2018})},\ \Eprint
  {https://arxiv.org/abs/1712.03398} {arXiv:1712.03398 [gr-qc]} \BibitemShut
  {NoStop}%
\bibitem [{\citenamefont {Witten}(1998)}]{Witten:1998qj}%
  \BibitemOpen
  \bibfield  {author} {\bibinfo {author} {\bibfnamefont {E.}~\bibnamefont
  {Witten}},\ }\bibfield  {title} {\bibinfo {title} {{Anti de Sitter space and
  holography}},\ }\href {https://doi.org/10.4310/ATMP.1998.v2.n2.a2} {\bibfield
   {journal} {\bibinfo  {journal} {Adv. Theor. Math. Phys.}\ }\textbf {\bibinfo
  {volume} {2}},\ \bibinfo {pages} {253} (\bibinfo {year} {1998})},\ \Eprint
  {https://arxiv.org/abs/hep-th/9802150} {arXiv:hep-th/9802150} \BibitemShut
  {NoStop}%
\bibitem [{\citenamefont {Maldacena}(1998)}]{Maldacena:1997re}%
  \BibitemOpen
  \bibfield  {author} {\bibinfo {author} {\bibfnamefont {J.~M.}\ \bibnamefont
  {Maldacena}},\ }\bibfield  {title} {\bibinfo {title} {{The Large $N$ limit of
  superconformal field theories and supergravity}},\ }\href
  {https://doi.org/10.4310/ATMP.1998.v2.n2.a1} {\bibfield  {journal} {\bibinfo
  {journal} {Adv. Theor. Math. Phys.}\ }\textbf {\bibinfo {volume} {2}},\
  \bibinfo {pages} {231} (\bibinfo {year} {1998})},\ \Eprint
  {https://arxiv.org/abs/hep-th/9711200} {arXiv:hep-th/9711200} \BibitemShut
  {NoStop}%
\bibitem [{\citenamefont {Bizon}\ and\ \citenamefont
  {Rostworowski}(2011)}]{Bizon:2011gg}%
  \BibitemOpen
  \bibfield  {author} {\bibinfo {author} {\bibfnamefont {P.}~\bibnamefont
  {Bizon}}\ and\ \bibinfo {author} {\bibfnamefont {A.}~\bibnamefont
  {Rostworowski}},\ }\bibfield  {title} {\bibinfo {title} {{On weakly turbulent
  instability of anti-de Sitter space}},\ }\href
  {https://doi.org/10.1103/PhysRevLett.107.031102} {\bibfield  {journal}
  {\bibinfo  {journal} {Phys. Rev. Lett.}\ }\textbf {\bibinfo {volume} {107}},\
  \bibinfo {pages} {031102} (\bibinfo {year} {2011})},\ \Eprint
  {https://arxiv.org/abs/1104.3702} {arXiv:1104.3702 [gr-qc]} \BibitemShut
  {NoStop}%
\bibitem [{\citenamefont {Ja\l{}mu\ifmmode~\dot{z}\else \.{z}\fi{}na}\ \emph
  {et~al.}(2011)\citenamefont {Ja\l{}mu\ifmmode~\dot{z}\else \.{z}\fi{}na},
  \citenamefont {Rostworowski},\ and\ \citenamefont
  {Bizo\ifmmode~\acute{n}\else \'{n}\fi{}}}]{PhysRevD.84.085021}%
  \BibitemOpen
  \bibfield  {author} {\bibinfo {author} {\bibfnamefont {J.}~\bibnamefont
  {Ja\l{}mu\ifmmode~\dot{z}\else \.{z}\fi{}na}}, \bibinfo {author}
  {\bibfnamefont {A.}~\bibnamefont {Rostworowski}},\ and\ \bibinfo {author}
  {\bibfnamefont {P.}~\bibnamefont {Bizo\ifmmode~\acute{n}\else \'{n}\fi{}}},\
  }\bibfield  {title} {\bibinfo {title} {Ads collapse of a scalar field in
  higher dimensions},\ }\href {https://doi.org/10.1103/PhysRevD.84.085021}
  {\bibfield  {journal} {\bibinfo  {journal} {Phys. Rev. D}\ }\textbf {\bibinfo
  {volume} {84}},\ \bibinfo {pages} {085021} (\bibinfo {year}
  {2011})}\BibitemShut {NoStop}%
\bibitem [{\citenamefont {Buchel}\ \emph {et~al.}(2012)\citenamefont {Buchel},
  \citenamefont {Lehner},\ and\ \citenamefont {Liebling}}]{Buchel:2012uh}%
  \BibitemOpen
  \bibfield  {author} {\bibinfo {author} {\bibfnamefont {A.}~\bibnamefont
  {Buchel}}, \bibinfo {author} {\bibfnamefont {L.}~\bibnamefont {Lehner}},\
  and\ \bibinfo {author} {\bibfnamefont {S.~L.}\ \bibnamefont {Liebling}},\
  }\bibfield  {title} {\bibinfo {title} {{Scalar Collapse in AdS}},\ }\href
  {https://doi.org/10.1103/PhysRevD.86.123011} {\bibfield  {journal} {\bibinfo
  {journal} {Phys. Rev. D}\ }\textbf {\bibinfo {volume} {86}},\ \bibinfo
  {pages} {123011} (\bibinfo {year} {2012})},\ \Eprint
  {https://arxiv.org/abs/1210.0890} {arXiv:1210.0890 [gr-qc]} \BibitemShut
  {NoStop}%
\bibitem [{\citenamefont {Dias}\ \emph {et~al.}(2012)\citenamefont {Dias},
  \citenamefont {Horowitz}, \citenamefont {Marolf},\ and\ \citenamefont
  {Santos}}]{Dias:2012tq}%
  \BibitemOpen
  \bibfield  {author} {\bibinfo {author} {\bibfnamefont {O.~J.~C.}\
  \bibnamefont {Dias}}, \bibinfo {author} {\bibfnamefont {G.~T.}\ \bibnamefont
  {Horowitz}}, \bibinfo {author} {\bibfnamefont {D.}~\bibnamefont {Marolf}},\
  and\ \bibinfo {author} {\bibfnamefont {J.~E.}\ \bibnamefont {Santos}},\
  }\bibfield  {title} {\bibinfo {title} {{On the Nonlinear Stability of
  Asymptotically Anti-de Sitter Solutions}},\ }\href
  {https://doi.org/10.1088/0264-9381/29/23/235019} {\bibfield  {journal}
  {\bibinfo  {journal} {Class. Quant. Grav.}\ }\textbf {\bibinfo {volume}
  {29}},\ \bibinfo {pages} {235019} (\bibinfo {year} {2012})},\ \Eprint
  {https://arxiv.org/abs/1208.5772} {arXiv:1208.5772 [gr-qc]} \BibitemShut
  {NoStop}%
\bibitem [{\citenamefont {Alcubierre}\ \emph {et~al.}(2018)\citenamefont
  {Alcubierre}, \citenamefont {Barranco}, \citenamefont {Bernal}, \citenamefont
  {Degollado}, \citenamefont {Diez-Tejedor}, \citenamefont {Megevand},
  \citenamefont {Nunez},\ and\ \citenamefont {Sarbach}}]{Alcubierre:2018ahf}%
  \BibitemOpen
  \bibfield  {author} {\bibinfo {author} {\bibfnamefont {M.}~\bibnamefont
  {Alcubierre}}, \bibinfo {author} {\bibfnamefont {J.}~\bibnamefont
  {Barranco}}, \bibinfo {author} {\bibfnamefont {A.}~\bibnamefont {Bernal}},
  \bibinfo {author} {\bibfnamefont {J.~C.}\ \bibnamefont {Degollado}}, \bibinfo
  {author} {\bibfnamefont {A.}~\bibnamefont {Diez-Tejedor}}, \bibinfo {author}
  {\bibfnamefont {M.}~\bibnamefont {Megevand}}, \bibinfo {author}
  {\bibfnamefont {D.}~\bibnamefont {Nunez}},\ and\ \bibinfo {author}
  {\bibfnamefont {O.}~\bibnamefont {Sarbach}},\ }\bibfield  {title} {\bibinfo
  {title} {{$\ell$-Boson stars}},\ }\href
  {https://doi.org/10.1088/1361-6382/aadcb6} {\bibfield  {journal} {\bibinfo
  {journal} {Class. Quant. Grav.}\ }\textbf {\bibinfo {volume} {35}},\ \bibinfo
  {pages} {19LT01} (\bibinfo {year} {2018})},\ \Eprint
  {https://arxiv.org/abs/1805.11488} {arXiv:1805.11488 [gr-qc]} \BibitemShut
  {NoStop}%
\bibitem [{\citenamefont {Alcubierre}\ \emph {et~al.}(2023)\citenamefont
  {Alcubierre}, \citenamefont {Barranco}, \citenamefont {Bernal}, \citenamefont
  {Degollado}, \citenamefont {Diez-Tejedor}, \citenamefont {Megevand},
  \citenamefont {N{\'u}{\~n}ez},\ and\ \citenamefont
  {Sarbach}}]{Alcubierre:2022rgp}%
  \BibitemOpen
  \bibfield  {author} {\bibinfo {author} {\bibfnamefont {M.}~\bibnamefont
  {Alcubierre}}, \bibinfo {author} {\bibfnamefont {J.}~\bibnamefont
  {Barranco}}, \bibinfo {author} {\bibfnamefont {A.}~\bibnamefont {Bernal}},
  \bibinfo {author} {\bibfnamefont {J.~C.}\ \bibnamefont {Degollado}}, \bibinfo
  {author} {\bibfnamefont {A.}~\bibnamefont {Diez-Tejedor}}, \bibinfo {author}
  {\bibfnamefont {M.}~\bibnamefont {Megevand}}, \bibinfo {author}
  {\bibfnamefont {D.}~\bibnamefont {N{\'u}{\~n}ez}},\ and\ \bibinfo {author}
  {\bibfnamefont {O.}~\bibnamefont {Sarbach}},\ }\bibfield  {title} {\bibinfo
  {title} {{Boson stars and their relatives in semiclassical gravity}},\ }\href
  {https://doi.org/10.1103/PhysRevD.107.045017} {\bibfield  {journal} {\bibinfo
   {journal} {Phys. Rev. D}\ }\textbf {\bibinfo {volume} {107}},\ \bibinfo
  {pages} {045017} (\bibinfo {year} {2023})},\ \Eprint
  {https://arxiv.org/abs/2212.02530} {arXiv:2212.02530 [gr-qc]} \BibitemShut
  {NoStop}%
\bibitem [{\citenamefont {Alcubierre}\ \emph {et~al.}(2022)\citenamefont
  {Alcubierre}, \citenamefont {Barranco}, \citenamefont {Bernal}, \citenamefont
  {Degollado}, \citenamefont {Diez-Tejedor}, \citenamefont {Jaramillo},
  \citenamefont {Megevand}, \citenamefont {N{\'u}{\~n}ez},\ and\ \citenamefont
  {Sarbach}}]{Alcubierre:2021psa}%
  \BibitemOpen
  \bibfield  {author} {\bibinfo {author} {\bibfnamefont {M.}~\bibnamefont
  {Alcubierre}}, \bibinfo {author} {\bibfnamefont {J.}~\bibnamefont
  {Barranco}}, \bibinfo {author} {\bibfnamefont {A.}~\bibnamefont {Bernal}},
  \bibinfo {author} {\bibfnamefont {J.~C.}\ \bibnamefont {Degollado}}, \bibinfo
  {author} {\bibfnamefont {A.}~\bibnamefont {Diez-Tejedor}}, \bibinfo {author}
  {\bibfnamefont {V.}~\bibnamefont {Jaramillo}}, \bibinfo {author}
  {\bibfnamefont {M.}~\bibnamefont {Megevand}}, \bibinfo {author}
  {\bibfnamefont {D.}~\bibnamefont {N{\'u}{\~n}ez}},\ and\ \bibinfo {author}
  {\bibfnamefont {O.}~\bibnamefont {Sarbach}},\ }\bibfield  {title} {\bibinfo
  {title} {{Extreme {\ensuremath{\ell}}-boson stars}},\ }\href
  {https://doi.org/10.1088/1361-6382/ac5fc2} {\bibfield  {journal} {\bibinfo
  {journal} {Class. Quant. Grav.}\ }\textbf {\bibinfo {volume} {39}},\ \bibinfo
  {pages} {094001} (\bibinfo {year} {2022})},\ \Eprint
  {https://arxiv.org/abs/2112.04529} {arXiv:2112.04529 [gr-qc]} \BibitemShut
  {NoStop}%
\bibitem [{\citenamefont {Alcubierre}\ \emph {et~al.}(2021)\citenamefont
  {Alcubierre}, \citenamefont {Barranco}, \citenamefont {Bernal}, \citenamefont
  {Degollado}, \citenamefont {Diez-Tejedor}, \citenamefont {Megevand},
  \citenamefont {N{\'u}{\~n}ez},\ and\ \citenamefont
  {Sarbach}}]{Alcubierre:2021mvs}%
  \BibitemOpen
  \bibfield  {author} {\bibinfo {author} {\bibfnamefont {M.}~\bibnamefont
  {Alcubierre}}, \bibinfo {author} {\bibfnamefont {J.}~\bibnamefont
  {Barranco}}, \bibinfo {author} {\bibfnamefont {A.}~\bibnamefont {Bernal}},
  \bibinfo {author} {\bibfnamefont {J.~C.}\ \bibnamefont {Degollado}}, \bibinfo
  {author} {\bibfnamefont {A.}~\bibnamefont {Diez-Tejedor}}, \bibinfo {author}
  {\bibfnamefont {M.}~\bibnamefont {Megevand}}, \bibinfo {author}
  {\bibfnamefont {D.}~\bibnamefont {N{\'u}{\~n}ez}},\ and\ \bibinfo {author}
  {\bibfnamefont {O.}~\bibnamefont {Sarbach}},\ }\bibfield  {title} {\bibinfo
  {title} {{On the linear stability of {\ensuremath{\ell}}-boson stars with
  respect to radial perturbations}},\ }\href
  {https://doi.org/10.1088/1361-6382/ac0160} {\bibfield  {journal} {\bibinfo
  {journal} {Class. Quant. Grav.}\ }\textbf {\bibinfo {volume} {38}},\ \bibinfo
  {pages} {174001} (\bibinfo {year} {2021})},\ \Eprint
  {https://arxiv.org/abs/2103.15012} {arXiv:2103.15012 [gr-qc]} \BibitemShut
  {NoStop}%
\bibitem [{\citenamefont {Alcubierre}\ \emph {et~al.}(2019)\citenamefont
  {Alcubierre}, \citenamefont {Barranco}, \citenamefont {Bernal}, \citenamefont
  {Degollado}, \citenamefont {Diez-Tejedor}, \citenamefont {Megevand},
  \citenamefont {N{\'u}{\~n}ez},\ and\ \citenamefont
  {Sarbach}}]{Alcubierre:2019qnh}%
  \BibitemOpen
  \bibfield  {author} {\bibinfo {author} {\bibfnamefont {M.}~\bibnamefont
  {Alcubierre}}, \bibinfo {author} {\bibfnamefont {J.}~\bibnamefont
  {Barranco}}, \bibinfo {author} {\bibfnamefont {A.}~\bibnamefont {Bernal}},
  \bibinfo {author} {\bibfnamefont {J.~C.}\ \bibnamefont {Degollado}}, \bibinfo
  {author} {\bibfnamefont {A.}~\bibnamefont {Diez-Tejedor}}, \bibinfo {author}
  {\bibfnamefont {M.}~\bibnamefont {Megevand}}, \bibinfo {author}
  {\bibfnamefont {D.}~\bibnamefont {N{\'u}{\~n}ez}},\ and\ \bibinfo {author}
  {\bibfnamefont {O.}~\bibnamefont {Sarbach}},\ }\bibfield  {title} {\bibinfo
  {title} {{Dynamical evolutions of $\ell$-boson stars in spherical
  symmetry}},\ }\href {https://doi.org/10.1088/1361-6382/ab4726} {\bibfield
  {journal} {\bibinfo  {journal} {Class. Quant. Grav.}\ }\textbf {\bibinfo
  {volume} {36}},\ \bibinfo {pages} {215013} (\bibinfo {year} {2019})},\
  \Eprint {https://arxiv.org/abs/1906.08959} {arXiv:1906.08959 [gr-qc]}
  \BibitemShut {NoStop}%
\bibitem [{\citenamefont {Jaramillo}\ \emph {et~al.}(2020)\citenamefont
  {Jaramillo}, \citenamefont {Sanchis-Gual}, \citenamefont {Barranco},
  \citenamefont {Bernal}, \citenamefont {Degollado}, \citenamefont {Herdeiro},\
  and\ \citenamefont {N{\'u}{\~n}ez}}]{Jaramillo:2020rsv}%
  \BibitemOpen
  \bibfield  {author} {\bibinfo {author} {\bibfnamefont {V.}~\bibnamefont
  {Jaramillo}}, \bibinfo {author} {\bibfnamefont {N.}~\bibnamefont
  {Sanchis-Gual}}, \bibinfo {author} {\bibfnamefont {J.}~\bibnamefont
  {Barranco}}, \bibinfo {author} {\bibfnamefont {A.}~\bibnamefont {Bernal}},
  \bibinfo {author} {\bibfnamefont {J.~C.}\ \bibnamefont {Degollado}}, \bibinfo
  {author} {\bibfnamefont {C.}~\bibnamefont {Herdeiro}},\ and\ \bibinfo
  {author} {\bibfnamefont {D.}~\bibnamefont {N{\'u}{\~n}ez}},\ }\bibfield
  {title} {\bibinfo {title} {{Dynamical {\ensuremath{\ell}} -boson stars:
  Generic stability and evidence for nonspherical solutions}},\ }\href
  {https://doi.org/10.1103/PhysRevD.101.124020} {\bibfield  {journal} {\bibinfo
   {journal} {Phys. Rev. D}\ }\textbf {\bibinfo {volume} {101}},\ \bibinfo
  {pages} {124020} (\bibinfo {year} {2020})},\ \Eprint
  {https://arxiv.org/abs/2004.08459} {arXiv:2004.08459 [gr-qc]} \BibitemShut
  {NoStop}%
\bibitem [{\citenamefont {Jaramillo}\ \emph {et~al.}(2022)\citenamefont
  {Jaramillo}, \citenamefont {Sanchis-Gual}, \citenamefont {Barranco},
  \citenamefont {Bernal}, \citenamefont {Degollado}, \citenamefont {Herdeiro},
  \citenamefont {Megevand},\ and\ \citenamefont
  {N{\'u}{\~n}ez}}]{Jaramillo:2022zwg}%
  \BibitemOpen
  \bibfield  {author} {\bibinfo {author} {\bibfnamefont {V.}~\bibnamefont
  {Jaramillo}}, \bibinfo {author} {\bibfnamefont {N.}~\bibnamefont
  {Sanchis-Gual}}, \bibinfo {author} {\bibfnamefont {J.}~\bibnamefont
  {Barranco}}, \bibinfo {author} {\bibfnamefont {A.}~\bibnamefont {Bernal}},
  \bibinfo {author} {\bibfnamefont {J.~C.}\ \bibnamefont {Degollado}}, \bibinfo
  {author} {\bibfnamefont {C.}~\bibnamefont {Herdeiro}}, \bibinfo {author}
  {\bibfnamefont {M.}~\bibnamefont {Megevand}},\ and\ \bibinfo {author}
  {\bibfnamefont {D.}~\bibnamefont {N{\'u}{\~n}ez}},\ }\bibfield  {title}
  {\bibinfo {title} {{Head-on collisions of {\ensuremath{\ell}}-boson stars}},\
  }\href {https://doi.org/10.1103/PhysRevD.105.104057} {\bibfield  {journal}
  {\bibinfo  {journal} {Phys. Rev. D}\ }\textbf {\bibinfo {volume} {105}},\
  \bibinfo {pages} {104057} (\bibinfo {year} {2022})},\ \Eprint
  {https://arxiv.org/abs/2202.00696} {arXiv:2202.00696 [gr-qc]} \BibitemShut
  {NoStop}%
\bibitem [{\citenamefont {Roque}\ \emph {et~al.}(2023)\citenamefont {Roque},
  \citenamefont {Nambo},\ and\ \citenamefont {Sarbach}}]{Roque:2023sjl}%
  \BibitemOpen
  \bibfield  {author} {\bibinfo {author} {\bibfnamefont {A.~A.}\ \bibnamefont
  {Roque}}, \bibinfo {author} {\bibfnamefont {E.~C.}\ \bibnamefont {Nambo}},\
  and\ \bibinfo {author} {\bibfnamefont {O.}~\bibnamefont {Sarbach}},\
  }\bibfield  {title} {\bibinfo {title} {{Radial linear stability of
  nonrelativistic {\ensuremath{\ell}}-boson stars}},\ }\href
  {https://doi.org/10.1103/PhysRevD.107.084001} {\bibfield  {journal} {\bibinfo
   {journal} {Phys. Rev. D}\ }\textbf {\bibinfo {volume} {107}},\ \bibinfo
  {pages} {084001} (\bibinfo {year} {2023})},\ \Eprint
  {https://arxiv.org/abs/2302.00717} {arXiv:2302.00717 [gr-qc]} \BibitemShut
  {NoStop}%
\bibitem [{\citenamefont {Nambo}\ \emph {et~al.}(2023)\citenamefont {Nambo},
  \citenamefont {Roque},\ and\ \citenamefont {Sarbach}}]{Nambo:2023yut}%
  \BibitemOpen
  \bibfield  {author} {\bibinfo {author} {\bibfnamefont {E.~C.}\ \bibnamefont
  {Nambo}}, \bibinfo {author} {\bibfnamefont {A.~A.}\ \bibnamefont {Roque}},\
  and\ \bibinfo {author} {\bibfnamefont {O.}~\bibnamefont {Sarbach}},\
  }\bibfield  {title} {\bibinfo {title} {{Are nonrelativistic ground state
  {\ensuremath{\ell}}-boson stars only stable for {\ensuremath{\ell}}=0 and
  {\ensuremath{\ell}}=1?}},\ }\href
  {https://doi.org/10.1103/PhysRevD.108.124065} {\bibfield  {journal} {\bibinfo
   {journal} {Phys. Rev. D}\ }\textbf {\bibinfo {volume} {108}},\ \bibinfo
  {pages} {124065} (\bibinfo {year} {2023})},\ \Eprint
  {https://arxiv.org/abs/2310.18405} {arXiv:2310.18405 [gr-qc]} \BibitemShut
  {NoStop}%
\bibitem [{\citenamefont {Wald}(1984)}]{wald_general_1984}%
  \BibitemOpen
  \bibfield  {author} {\bibinfo {author} {\bibfnamefont {R.~M.}\ \bibnamefont
  {Wald}},\ }\href {https://doi.org/10.7208/chicago/9780226870373.001.0001}
  {\emph {\bibinfo {title} {General Relativity}}}\ (\bibinfo  {publisher}
  {University of Chicago Press},\ \bibinfo {address} {Chicago, U.S.A.},\
  \bibinfo {year} {1984})\BibitemShut {NoStop}%
\bibitem [{\citenamefont {Cunha}\ \emph {et~al.}(2017)\citenamefont {Cunha},
  \citenamefont {Berti},\ and\ \citenamefont {Herdeiro}}]{Cunha:2017qtt}%
  \BibitemOpen
  \bibfield  {author} {\bibinfo {author} {\bibfnamefont {P.~V.~P.}\
  \bibnamefont {Cunha}}, \bibinfo {author} {\bibfnamefont {E.}~\bibnamefont
  {Berti}},\ and\ \bibinfo {author} {\bibfnamefont {C.~A.~R.}\ \bibnamefont
  {Herdeiro}},\ }\bibfield  {title} {\bibinfo {title} {{Light-Ring Stability
  for Ultracompact Objects}},\ }\href
  {https://doi.org/10.1103/PhysRevLett.119.251102} {\bibfield  {journal}
  {\bibinfo  {journal} {Phys. Rev. Lett.}\ }\textbf {\bibinfo {volume} {119}},\
  \bibinfo {pages} {251102} (\bibinfo {year} {2017})},\ \Eprint
  {https://arxiv.org/abs/1708.04211} {arXiv:1708.04211 [gr-qc]} \BibitemShut
  {NoStop}%
\bibitem [{\citenamefont {Guo}\ \emph {et~al.}(2022)\citenamefont {Guo},
  \citenamefont {Zhong}, \citenamefont {Wang},\ and\ \citenamefont
  {Gao}}]{Guo:2021bcw}%
  \BibitemOpen
  \bibfield  {author} {\bibinfo {author} {\bibfnamefont {M.}~\bibnamefont
  {Guo}}, \bibinfo {author} {\bibfnamefont {Z.}~\bibnamefont {Zhong}}, \bibinfo
  {author} {\bibfnamefont {J.}~\bibnamefont {Wang}},\ and\ \bibinfo {author}
  {\bibfnamefont {S.}~\bibnamefont {Gao}},\ }\bibfield  {title} {\bibinfo
  {title} {{Light rings and long-lived modes in quasiblack hole spacetimes}},\
  }\href {https://doi.org/10.1103/PhysRevD.105.024049} {\bibfield  {journal}
  {\bibinfo  {journal} {Phys. Rev. D}\ }\textbf {\bibinfo {volume} {105}},\
  \bibinfo {pages} {024049} (\bibinfo {year} {2022})},\ \Eprint
  {https://arxiv.org/abs/2108.08967} {arXiv:2108.08967 [gr-qc]} \BibitemShut
  {NoStop}%
\bibitem [{\citenamefont {Peng}(2020)}]{Peng:2019uzw}%
  \BibitemOpen
  \bibfield  {author} {\bibinfo {author} {\bibfnamefont {Y.}~\bibnamefont
  {Peng}},\ }\bibfield  {title} {\bibinfo {title} {{No-go theorem for static
  boson stars with negative cosmological constants}},\ }\href
  {https://doi.org/10.1016/j.nuclphysb.2020.114955} {\bibfield  {journal}
  {\bibinfo  {journal} {Nucl. Phys. B}\ }\textbf {\bibinfo {volume} {953}},\
  \bibinfo {pages} {114955} (\bibinfo {year} {2020})},\ \Eprint
  {https://arxiv.org/abs/1902.06508} {arXiv:1902.06508 [gr-qc]} \BibitemShut
  {NoStop}%
\bibitem [{\citenamefont {Press}\ \emph {et~al.}(1992)\citenamefont {Press},
  \citenamefont {Teukolsky}, \citenamefont {Vetterling},\ and\ \citenamefont
  {Flannery}}]{Press1992Numerical}%
  \BibitemOpen
  \bibfield  {author} {\bibinfo {author} {\bibfnamefont {W.~H.}\ \bibnamefont
  {Press}}, \bibinfo {author} {\bibfnamefont {S.~A.}\ \bibnamefont
  {Teukolsky}}, \bibinfo {author} {\bibfnamefont {W.~T.}\ \bibnamefont
  {Vetterling}},\ and\ \bibinfo {author} {\bibfnamefont {B.~P.}\ \bibnamefont
  {Flannery}},\ }\href {http://www.nr.com} {\emph {\bibinfo {title} {Numerical
  Recipes in FORTRAN: The Art of Scientific Computing}}},\ \bibinfo {edition}
  {2nd}\ ed.\ (\bibinfo  {publisher} {Cambridge University Press},\ \bibinfo
  {address} {Cambridge},\ \bibinfo {year} {1992})\BibitemShut {NoStop}%
\bibitem [{\citenamefont {Radhakrishnan}\ and\ \citenamefont
  {Hindmarsh}(1993)}]{Radhakrishnan1993}%
  \BibitemOpen
  \bibfield  {author} {\bibinfo {author} {\bibfnamefont {K.}~\bibnamefont
  {Radhakrishnan}}\ and\ \bibinfo {author} {\bibfnamefont {A.~C.}\ \bibnamefont
  {Hindmarsh}},\ }\href@noop {} {\emph {\bibinfo {title} {Description and use
  of LSODE, the Livermore solver for ordinary differential equations}}},\
  \bibinfo {type} {LLNL report}\ \bibinfo {number} {UCRL-ID-113855}\ (\bibinfo
  {institution} {Lawrence Livermore National Laboratory},\ \bibinfo {year}
  {1993})\BibitemShut {NoStop}%
\bibitem [{\citenamefont {Megevand}\ \emph {et~al.}(2007)\citenamefont
  {Megevand}, \citenamefont {Olabarrieta},\ and\ \citenamefont
  {Lehner}}]{Megevand:2007uy}%
  \BibitemOpen
  \bibfield  {author} {\bibinfo {author} {\bibfnamefont {M.}~\bibnamefont
  {Megevand}}, \bibinfo {author} {\bibfnamefont {I.}~\bibnamefont
  {Olabarrieta}},\ and\ \bibinfo {author} {\bibfnamefont {L.}~\bibnamefont
  {Lehner}},\ }\bibfield  {title} {\bibinfo {title} {{Scalar field confinement
  as a model for accreting systems}},\ }\href
  {https://doi.org/10.1088/0264-9381/24/13/007} {\bibfield  {journal} {\bibinfo
   {journal} {Class. Quant. Grav.}\ }\textbf {\bibinfo {volume} {24}},\
  \bibinfo {pages} {3235} (\bibinfo {year} {2007})},\ \Eprint
  {https://arxiv.org/abs/0705.0644} {arXiv:0705.0644 [gr-qc]} \BibitemShut
  {NoStop}%
\bibitem [{\citenamefont {Balakrishna}\ \emph {et~al.}(1998)\citenamefont
  {Balakrishna}, \citenamefont {Seidel},\ and\ \citenamefont
  {Suen}}]{Balakrishna:1997ej}%
  \BibitemOpen
  \bibfield  {author} {\bibinfo {author} {\bibfnamefont {J.}~\bibnamefont
  {Balakrishna}}, \bibinfo {author} {\bibfnamefont {E.}~\bibnamefont
  {Seidel}},\ and\ \bibinfo {author} {\bibfnamefont {W.-M.}\ \bibnamefont
  {Suen}},\ }\bibfield  {title} {\bibinfo {title} {{Dynamical evolution of
  boson stars. 2. Excited states and selfinteracting fields}},\ }\href
  {https://doi.org/10.1103/PhysRevD.58.104004} {\bibfield  {journal} {\bibinfo
  {journal} {Phys. Rev. D}\ }\textbf {\bibinfo {volume} {58}},\ \bibinfo
  {pages} {104004} (\bibinfo {year} {1998})},\ \Eprint
  {https://arxiv.org/abs/gr-qc/9712064} {arXiv:gr-qc/9712064} \BibitemShut
  {NoStop}%
\bibitem [{\citenamefont {Bernal}\ \emph {et~al.}(2010)\citenamefont {Bernal},
  \citenamefont {Barranco}, \citenamefont {Alic},\ and\ \citenamefont
  {Palenzuela}}]{Bernal:2009zy}%
  \BibitemOpen
  \bibfield  {author} {\bibinfo {author} {\bibfnamefont {A.}~\bibnamefont
  {Bernal}}, \bibinfo {author} {\bibfnamefont {J.}~\bibnamefont {Barranco}},
  \bibinfo {author} {\bibfnamefont {D.}~\bibnamefont {Alic}},\ and\ \bibinfo
  {author} {\bibfnamefont {C.}~\bibnamefont {Palenzuela}},\ }\bibfield  {title}
  {\bibinfo {title} {{Multi-state Boson Stars}},\ }\href
  {https://doi.org/10.1103/PhysRevD.81.044031} {\bibfield  {journal} {\bibinfo
  {journal} {Phys. Rev. D}\ }\textbf {\bibinfo {volume} {81}},\ \bibinfo
  {pages} {044031} (\bibinfo {year} {2010})},\ \Eprint
  {https://arxiv.org/abs/0908.2435} {arXiv:0908.2435 [gr-qc]} \BibitemShut
  {NoStop}%
\bibitem [{\citenamefont {Buchdahl}(1959)}]{Buchdahl:1959zz}%
  \BibitemOpen
  \bibfield  {author} {\bibinfo {author} {\bibfnamefont {H.~A.}\ \bibnamefont
  {Buchdahl}},\ }\bibfield  {title} {\bibinfo {title} {{General Relativistic
  Fluid Spheres}},\ }\href {https://doi.org/10.1103/PhysRev.116.1027}
  {\bibfield  {journal} {\bibinfo  {journal} {Phys. Rev.}\ }\textbf {\bibinfo
  {volume} {116}},\ \bibinfo {pages} {1027} (\bibinfo {year}
  {1959})}\BibitemShut {NoStop}%
\bibitem [{\citenamefont {Barranco}\ \emph {et~al.}(2021)\citenamefont
  {Barranco}, \citenamefont {Chagoya}, \citenamefont {Diez-Tejedor},
  \citenamefont {Niz},\ and\ \citenamefont {Roque}}]{Barranco:2021auj}%
  \BibitemOpen
  \bibfield  {author} {\bibinfo {author} {\bibfnamefont {J.}~\bibnamefont
  {Barranco}}, \bibinfo {author} {\bibfnamefont {J.}~\bibnamefont {Chagoya}},
  \bibinfo {author} {\bibfnamefont {A.}~\bibnamefont {Diez-Tejedor}}, \bibinfo
  {author} {\bibfnamefont {G.}~\bibnamefont {Niz}},\ and\ \bibinfo {author}
  {\bibfnamefont {A.~A.}\ \bibnamefont {Roque}},\ }\bibfield  {title} {\bibinfo
  {title} {{Horndeski stars}},\ }\href
  {https://doi.org/10.1088/1475-7516/2021/10/022} {\bibfield  {journal}
  {\bibinfo  {journal} {JCAP}\ }\textbf {\bibinfo {volume} {10}},\ \bibinfo
  {pages} {022}},\ \Eprint {https://arxiv.org/abs/2108.01679} {arXiv:2108.01679
  [gr-qc]} \BibitemShut {NoStop}%
\bibitem [{\citenamefont {Hashimoto}\ \emph {et~al.}(2023)\citenamefont
  {Hashimoto}, \citenamefont {Sugiura}, \citenamefont {Sugiyama},\ and\
  \citenamefont {Yoda}}]{Hashimoto:2023buz}%
  \BibitemOpen
  \bibfield  {author} {\bibinfo {author} {\bibfnamefont {K.}~\bibnamefont
  {Hashimoto}}, \bibinfo {author} {\bibfnamefont {K.}~\bibnamefont {Sugiura}},
  \bibinfo {author} {\bibfnamefont {K.}~\bibnamefont {Sugiyama}},\ and\
  \bibinfo {author} {\bibfnamefont {T.}~\bibnamefont {Yoda}},\ }\bibfield
  {title} {\bibinfo {title} {{Photon sphere and quasinormal modes in
  AdS/CFT}},\ }\href {https://doi.org/10.1007/JHEP10(2023)149} {\bibfield
  {journal} {\bibinfo  {journal} {JHEP}\ }\textbf {\bibinfo {volume} {10}},\
  \bibinfo {pages} {149}},\ \Eprint {https://arxiv.org/abs/2307.00237}
  {arXiv:2307.00237 [hep-th]} \BibitemShut {NoStop}%
\bibitem [{\citenamefont {Riojas}\ and\ \citenamefont
  {Sun}(2023)}]{Riojas:2023pew}%
  \BibitemOpen
  \bibfield  {author} {\bibinfo {author} {\bibfnamefont {M.}~\bibnamefont
  {Riojas}}\ and\ \bibinfo {author} {\bibfnamefont {H.-Y.}\ \bibnamefont
  {Sun}},\ }\bibfield  {title} {\bibinfo {title} {{The Photon Sphere and the
  AdS/CFT Correspondence}},\ }\href@noop {} {\  (\bibinfo {year} {2023})},\
  \Eprint {https://arxiv.org/abs/2307.06415} {arXiv:2307.06415 [hep-th]}
  \BibitemShut {NoStop}%
\end{thebibliography}

%


\end{document}